\documentclass[letterpaper]{article}
\usepackage[utf8]{inputenc}
\usepackage[T1]{fontenc}
\usepackage{amsmath}
\usepackage{graphicx}
\usepackage{textcomp}
\usepackage{authblk}
\usepackage{lipsum}
\usepackage{geometry}
\usepackage{hyperref}
\usepackage{cite}
\usepackage{blindtext}
\usepackage{microtype}
\usepackage{array}
\usepackage{booktabs}
\usepackage{xcolor}
\usepackage{amsmath}

\usepackage[colorinlistoftodos,prependcaption,textsize=tiny]{todonotes}
\providecommand{\correspondingauthor}{$^*$}

\usepackage{siunitx}
\usepackage{xcolor}

\DeclareSIUnit{\electronvolt}{eV}
\DeclareSIUnit{\millielectronvolt}{meV}
\DeclareMathAlphabet{\mathbb}{U}{msb}{m}{n}
\DeclareSIUnit{\angstrom}{\text{Å}} 

\title{Simulation of Self-Assembled Monolayers of Polyalanine $\alpha$-Helices: Development and Application of an Effective Potential for Film Structure Predictions}

\author[1]{Hadis Ghodrati Saeini}
\author[1]{Kevin Preis}
\author[1]{Thi Ngoc Ha Nguyen}
\author[1]{Christoph Tegenkamp}
\author[1]{Sibylle Gemming}
\author[1,2]{Jeffrey Kelling}
\author[3]{Florian Günther\correspondingauthor}

\affil[1]{Institute of Physics, Technische Universität Chemnitz, 09107 Chemnitz, Germany}
\affil[2]{Institute for Radiation Physics, Helmholtz-Zentrum Dresden - Rossendorf, Dresden, Germany}
\affil[3]{Departamento de Física, Universidade Estadual Paulista, Instituto de Geociências e Ciências Exatas, Rio Claro, Brazil}

\date{*Email: florian.gunther@unesp.br}

\begin{document} 
    \maketitle
\begin{abstract}
Self-assembled monolayers of polyalanine $\alpha$-helices exhibit distinct structural phases with implications for chiral-induced spin selectivity. We combine scanning tunneling microscopy and theoretical modeling to reveal how chiral composition governs supramolecular organization. Enantiopure systems form hexagonal lattices, while racemic mixtures organize into rectangular phases with stripe-like features. Our interaction potentials derived from density-functional based tight binding calculations show that opposite-handed helix pairs exhibit stronger binding and closer packing, explaining the denser racemic structures. Crucially, we demonstrate that the observed STM contrast arises from anti-parallel alignment of opposite-handed helices rather than physical height variations. These findings establish fundamental structure-property relationships for designing peptide-based spintronic materials.
\end{abstract}
\section*{Keywords}
self-assembled monolayers, polyalanine $\alpha$-helices, interfacial ordering, effective interaction model

\section{Introduction}

\begin{figure}[!b]
\centering
\includegraphics[width=0.9\textwidth]{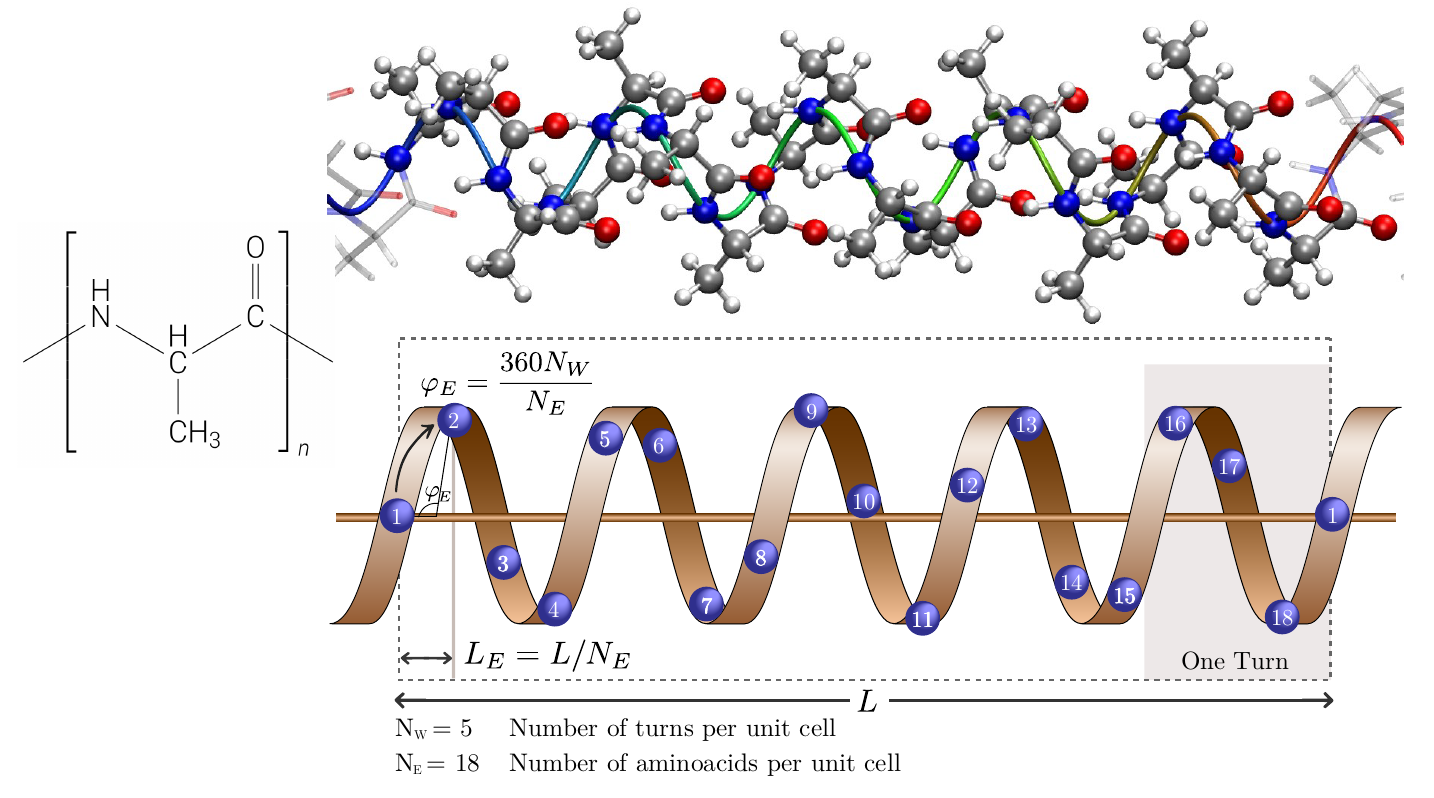}
\caption{Chemical structure of polyalanine (left) and schematic representation of a polyalanine $\alpha$-helix   (right). }
\label{fig:PA}
\end{figure}

Self-assembled monolayers (SAMs) play a fundamental role in modern materials science, with applications in nanotechnology, biosensing, and electronics.~\cite{mrksich2009using, venanzi2013self}. 
Among these, polypeptide SAMs are of significant interest due to their intrinsic ability to form chiral secondary structures, such as $\alpha$-helices, which can introduce advanced functionalities because they can mediate spin-selective transport, a phenomenon known as the chiral-induced spin selectivity (CISS) effect~\cite{Banerjee-Ghosh2018-px}.
 This effect enables efficient spin filtering without external magnetic fields or magnetic materials, thereby supporting advances in spintronics, where information is processed using electron spin rather than charge~\cite{bloom2024chiral, schmidt2025chiral, garcia2025ciss}. 
 Both experimental and theoretical work has demonstrated that helical polypeptides exhibit a pronounced CISS effect due to their rigid, chiral backbones, making them promising candidates for next-generation molecular spintronic devices~\cite{smith2020control, sek2006asymmetry, blondelle1997Polyalanine, Ghosh2020effect, sun2025unidirectional, nguyen2024mechanism}.

Despite these advances, fundamental questions regarding structure–property relationships in polypeptide SAMs remain open, particularly in the context of the CISS effect. 
A central issue is the mechanism of electronic transport through these systems, which lack the delocalized electronic structure common for organic conductors but yet exhibit highly efficient, spin-polarized charge transport~\cite{ha2019,ha2020}. 
Furthermore, the driving forces behind the self-assembly process are not fully understood.
Key questions include how the peptide sequence and specific functionalization influence the stability of different SAM phases and the kinetics of their formation, ultimately governing the supramolecular structure and its resulting electronic and spintronic properties~\cite{meier2023chiral, kulkarni2022spin, garcia2021chirality}.

Among the wide variety of polypeptides, the polyalanine $\alpha$-helix ($\alpha$PA, see Figure~\ref{fig:PA}) is particularly well-suited for studying self-assembly and the CISS effect, as it combines three key features:
First, it adopts an $\alpha$-helical structure, a common protein secondary structure stabilized by intra-helical hydrogen bonds~\cite{Kohtani2004,Hoffmann2016,ONeil1990}.
Second, its homopolypeptide nature, being composed solely of alanine residues, minimizes structural complexity and eliminates heterogeneous functionalization effects along the backbone.
Last, alanine is the smallest functionalized amino acid, with only a methyl group as a side chain, which results in a minimal set of parameters necessary to specify the configuration of the helix. 
Moreover, this small side chain also ensures that self-assembled, parallel-aligned $\alpha$PA molecules can pack densely due to reduced steric hindrance compared to polypeptides composed of larger amino acids.

\begin{figure}[!b]
\centering
\includegraphics[width=0.7\textwidth]{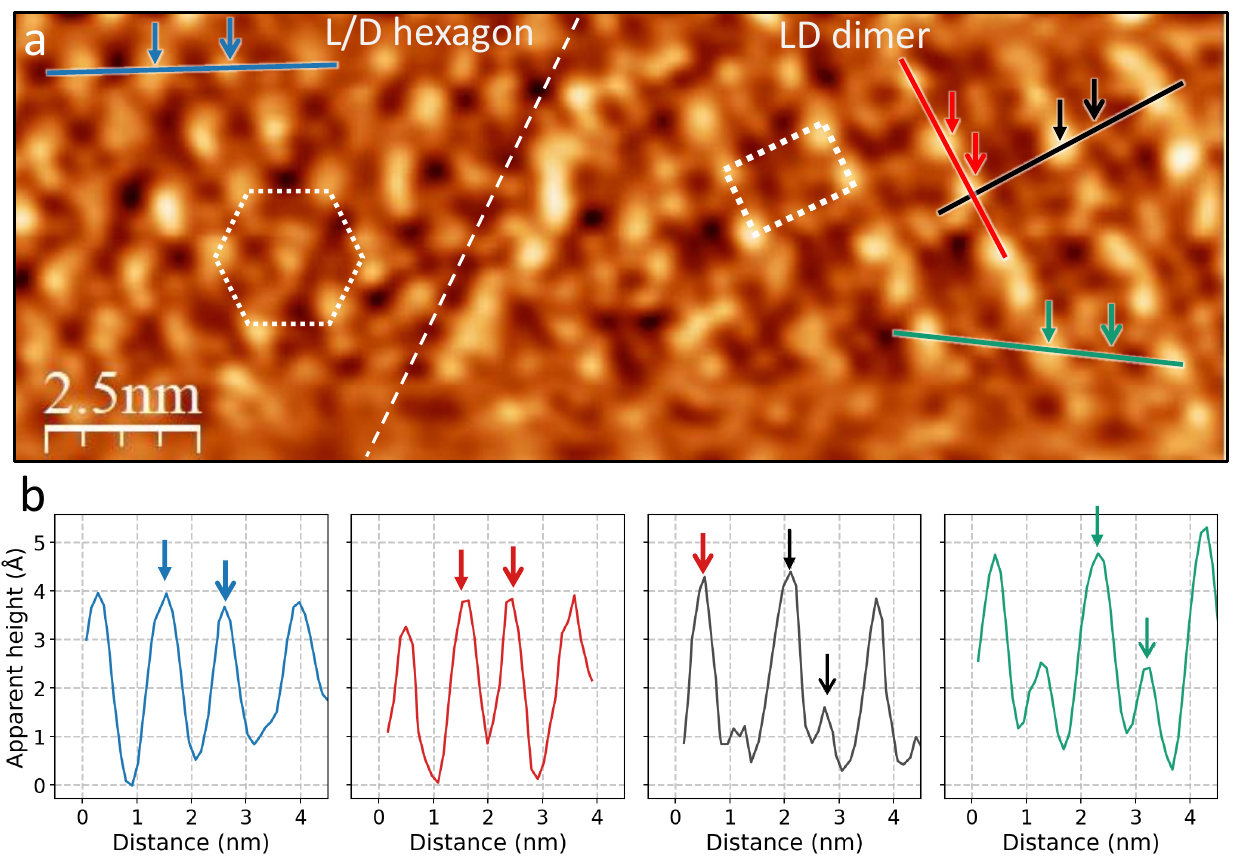}
\caption{(a) STM image of a self-assembled film of racemic LD-$\alpha$PA (LD-PA) on HOPG, showing the hexagonal phase of enantiopure L/D-PA (left) and the dimer phase of LD-PA (right). (b) Height profile taken along the colored lines in (a), showing the regular spacing and equal apparent heights in the hexagonal phase (blue arrows, left), variations in spacings but equal apparent heights along the parallel row (red arrows, middle-left), and regular spacing but differences in apparent height of adjacent rows within the dimer phase (green and black arrows, right).}
\label{fig:STM}
\end{figure}

Recently, films of enantiopure right-handed $\alpha$PA (L-PA) and racemic mixtures of right- and left-handed $\alpha$PA  (DL-PA) molecules formed on highly ordered pyrolytic graphite (HOPG) as well as on Al$_2$O$_3$/Pt/Au/Co/Au substrates were investigated under ambient conditions~\cite{ha2019,ha2020,ha2023} 
In these works, both structural properties and spin-resolved electronic conduction properties were studied using scanning tunneling microscopy (STM)  and scanning tunneling spectroscopy (STS), respectively. 
The latter revealed  CISS magnetoresistance (CISS-MR) for all observed systems, confirming the spin-selectivity properties of $\alpha$PA SAMs. 
In particular, it was found that the CISS-MR reached approximately 75\% for chemisorbed hexagonal phases but dropped to around 50\% for other phases. 
Additionally, the CISS-MR of chemisorbed molecules on Al$_2$O$_3$/Pt/Au/Co/Au substrates was up to 10\% higher than that of physisorbed molecules on HOPG~\cite{ha2023}. Furthermore, the STS measurements confirmed that the $\alpha$-helical conformation was preserved in both enantiopure and racemic films, with nearly identical HOMO–LUMO gaps ($\approx \SI{3.4}{eV}$)~\cite{ha2019}.
The STM characterization of the different $\alpha$PA SAMs revealed that for enantiopure L-PA on HOPG the adsorption resulted in a hexagonally close-packed (hcp) arrangement.
In contrast, two distinct phases were observed for racemic DL-PA films. 
On the one hand, an hcp structure with identical characteristics to the enantiopure L-PA system was detected. (Figure~\ref{fig:STM}a, left).
On the other hand, a novel phase with a rectangular unit cell has formed (Figure~\ref{fig:STM}a, right) \cite{ha2020}. 
While STM height profiles revealed nearly uniform heights for the hcp structure  (Figure~\ref{fig:STM}, blue line) and along the shorter axis of the rectangular phase (Figure~\ref{fig:STM}, red line),  alternating apparent heights of about \SI{3.0}{\angstrom} along the longer axis were observed (Figure~\ref{fig:STM}, black line). 
This apparent height difference reduces to about \SI{2.5}{\angstrom} four line scans diagonally to the rectangular cell (Figure~\ref{fig:STM}, green line), suggesting that an $\alpha$PA Helix is positioned in the center of the cell.
Furthermore, it was found that these dimer phases exhibited a \~25\% higher packing density compared to the hcp structure, which comes along with a reduced distance of $\approx\SI{0.4}{nm}$ between adjacent STM maximum features.  
This observation is consistent with Wallach’s rule which states that racemates can form denser structures than enantiopure systems~\cite{Wallach1895}.
Similar phases and lattice parameters were found also for chemically adsorbed $\alpha$PA molecules on Al$_2$O$_3$/Pt/Au/Co/Au substrates~\cite{ha2023}, suggesting that the film properties are dominated by the intermolecular features rather than the substrate.

The experimental observation of closer packing in the racemic mixture was interpreted assuming that intermolecular hydrogen bonds stabilize the individual phases, particularly the dimer phase~\cite{ha2023}.
Although this reasoning provides a phenomenological understanding, a detailed atomistic picture of the intermolecular interactions that stabilize the various SAM phases is still lacking. 
For example, the specific nature of such intermolecular hydrogen bonds remains unclear. 
A deeper knowledge of such interactions, however, is crucial, since it might alter the electronic structure properties of $\alpha$PA in SAMs compared to an isolated helix. 
In particular, the effect on the local charge distributions within the structure is important, since it has been recently shown experimentally that the orientation of the dipole can flip the sign of the CISS-MR~\cite{ha2024}.
Hence, computational work addressing the structural properties  of $\alpha$PA or similar polypeptides at the molecular scale is of high value to better understand the individual interactions  between neighboring $\alpha$PA molecules in SAMs.
Therefore, the here presented work aims to complement the experimental findings with theoretical insights into the intermolecular interactions within self-assembled structures of $\alpha$PA. Our goal is to identify the specific interactions responsible for experimentally observed effects, such as the formation of hcp and rectangular phases and the molecular offsets within them. 

\section{Methodology}

\subsection{Simulation of interaction at atomic scale}
To assess the geometrical properties of both isolated helices and helix pairs, a description of the interactions between all particles is required. 
For this purpose, we selected the self-consistent charge Density Functional based Tight Binding (SCC-DFTB) method, an approximate Kohn-Sham scheme~\cite{elstner1998self}. 
In contrast to classical force fields, SCC-DFTB is less empirical and directly provides the electronic structure of the system. 
While this electronic structure information is not the primary focus of the present work, it is crucial for future explorations of local dipole moments along the backbone, electronic transport properties,  and the assessment of the CISS effect. 
The SCC-DFTB computations were performed using the DFTB+ software (version 20.1) with the \texttt{mio-1-1} parameter set~\cite{dftbplus2020}. 
Moreover,  dispersion  was included using Grimme's dispersion correction using the universal force field parameters~\cite{grimme2011}.
While full geometry optimization was performed for isolated $\alpha$-PA helices, single-point energy calculations were conducted for helix pairs.

In all computations, the helices were modeled as ideal, infinite structures using periodic boundary conditions along the helical axis. 
Neglecting the termination groups is a valid approximation in cases where helices are physisorbed  so that inter-molecular interaction dominates the formation of SAM structures, such as $\alpha$PA on HOPG.
Moreover, this approach suppresses termination effects such as the interaction of global dipole moments, and thus limits the interaction to local intermolecular interactions.
Instead, considering infinite systems incorporates additional symmetry and, hence, reduces the degrees of freedom to a limited number of parameters, enabling a more sophisticated analysis of local interactions.
It is important to note, however, that for helix pairs or ensembles the use of periodic boundary conditions restricts the configurations to parallel or anti-parallel alignments. 

\subsection{\texorpdfstring{Symmetry of Isolated $\alpha$-PA Helices}{Symmetry of Isolated α-PA Helices}}
In proteins and polypeptides, the $\alpha$-helix is a well-known secondary structure, where the polypeptide backbone forms a helix stabilized by hydrogen bonds between the carbonyl group of one amino acid and the amide hydrogen of another one that is four residues away~\cite{Kohtani2004,Hoffmann2016}. 
This results in a structure in which each amino acid residue corresponds to a turn of $\phi_E = \ang{100}$ and a translation of $L_E \approx \SI{1.49}\angstrom$~along the helical axis. 
Hence, 3.6 amino acids form  a full turn with a pitch of approximately \SI{5.4}\angstrom~\cite{ONeil1990}. 
In an $\alpha$-helix structure the functional groups of the individual amino acids face outward and therefore dominate the interaction with the environment.

To model this helical structure, we consider a unit cell of 18 alanine residues over five turns, with a total length along the helix direction of $L \approx \SI{26.8}\angstrom$~(Figure~\ref{fig:PA}).
That way, the unit cell exhibits redundancy, as a simultaneous rotation by $\phi_E$ and translation by $L_E$ maps the molecular structure onto itself.

\subsection{Symmetry Properties of Helix Pairs}
To study the interaction between isolated helix pairs, a set of variables that uniquely defines all possible pair configurations must be chosen. 
To sample the interaction potential as densely as possible while conserving computational resources, it is important to avoid redundancy in the parameter space (i.e., not storing two configurations that represent the same atom arrangement due to symmetry). 
It is therefore meaningful to make use of the symmetry aspects to reduce the parameter space of the pair interaction.

While for a single isolated helix the absolute direction or handedness is unimportant, for helix pairs the relative alignment of the second helix with respect to the first must be considered. 
In addition to the helices' handedness (right- or left-handed), the relative orientation of molecular features, such as the direction of the carbonyl groups, impacts the inter-helical interaction.
Consequently, four distinct alignments must be considered, which are summarized in Table~\ref{tab:nomenclature}: (1) equally handed, parallel alignment (EP), (2) equally handed, anti-parallel alignment (EA), (3) oppositely handed, parallel alignment (OP), and (4) oppositely handed, anti-parallel alignment (OA).

\begin{table}[!htbp]
    \centering
    \caption{Nomenclature for helix pairs and multi-helix ensembles (SAM).}
    \label{tab:nomenclature}
    \renewcommand{\arraystretch}{1.3}
    \begin{tabular}{m{2cm} m{4.5cm} m{2cm}}
        \toprule
        \textbf{Short code} & \textbf{Description} & \textbf{Schematic} \\
        \midrule
        \multicolumn{3}{l}{\textit{Pair of helices}} \\
        \midrule
        EP &
        Equally handed,\newline
        aligned in parallel
        &
        \makebox[2cm][c]{\includegraphics[width=1.5cm]{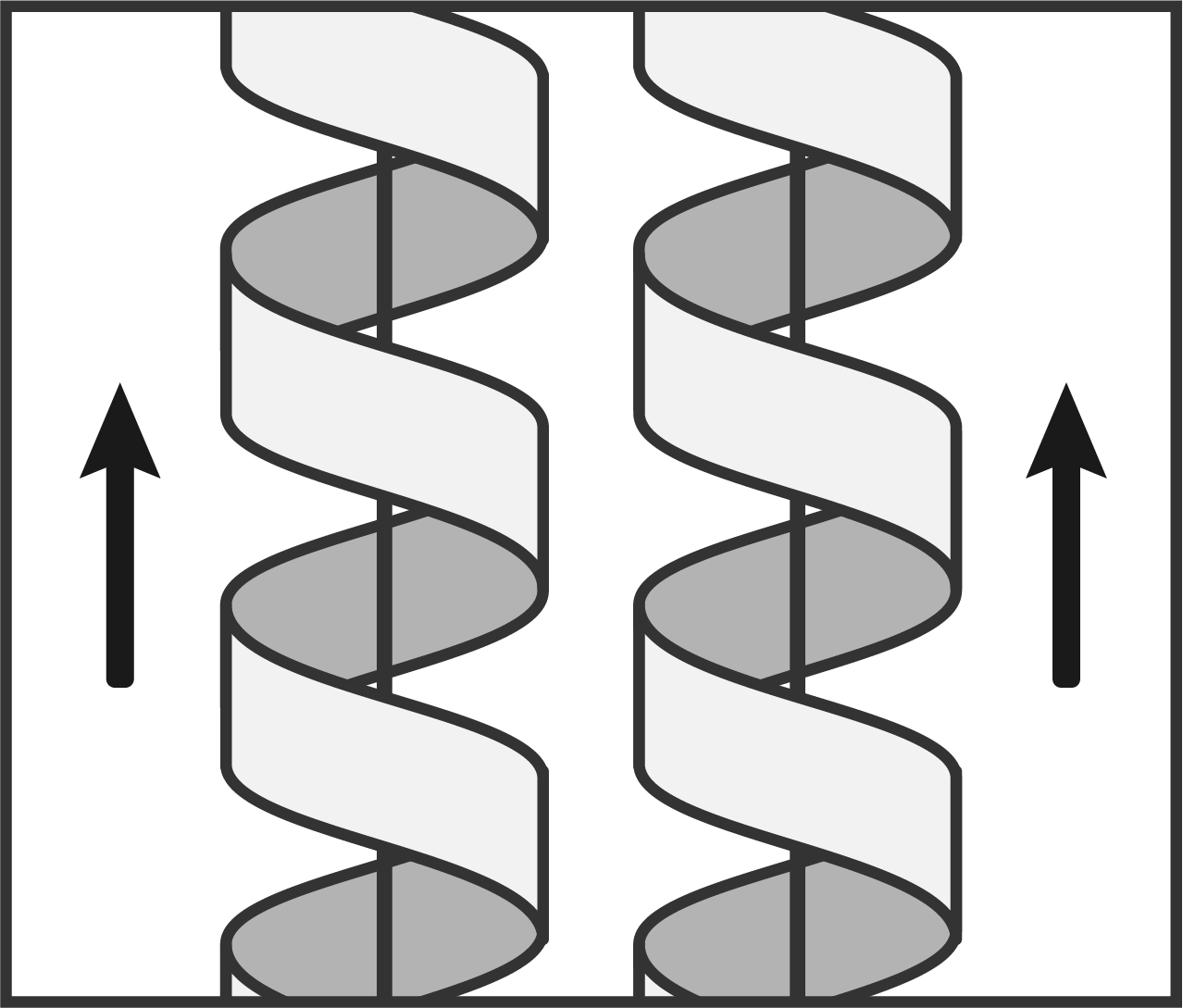}} \\
        EA &
        Equally handed,\newline
        aligned anti-parallel
        &
        \makebox[2cm][c]{\includegraphics[width=1.5cm]{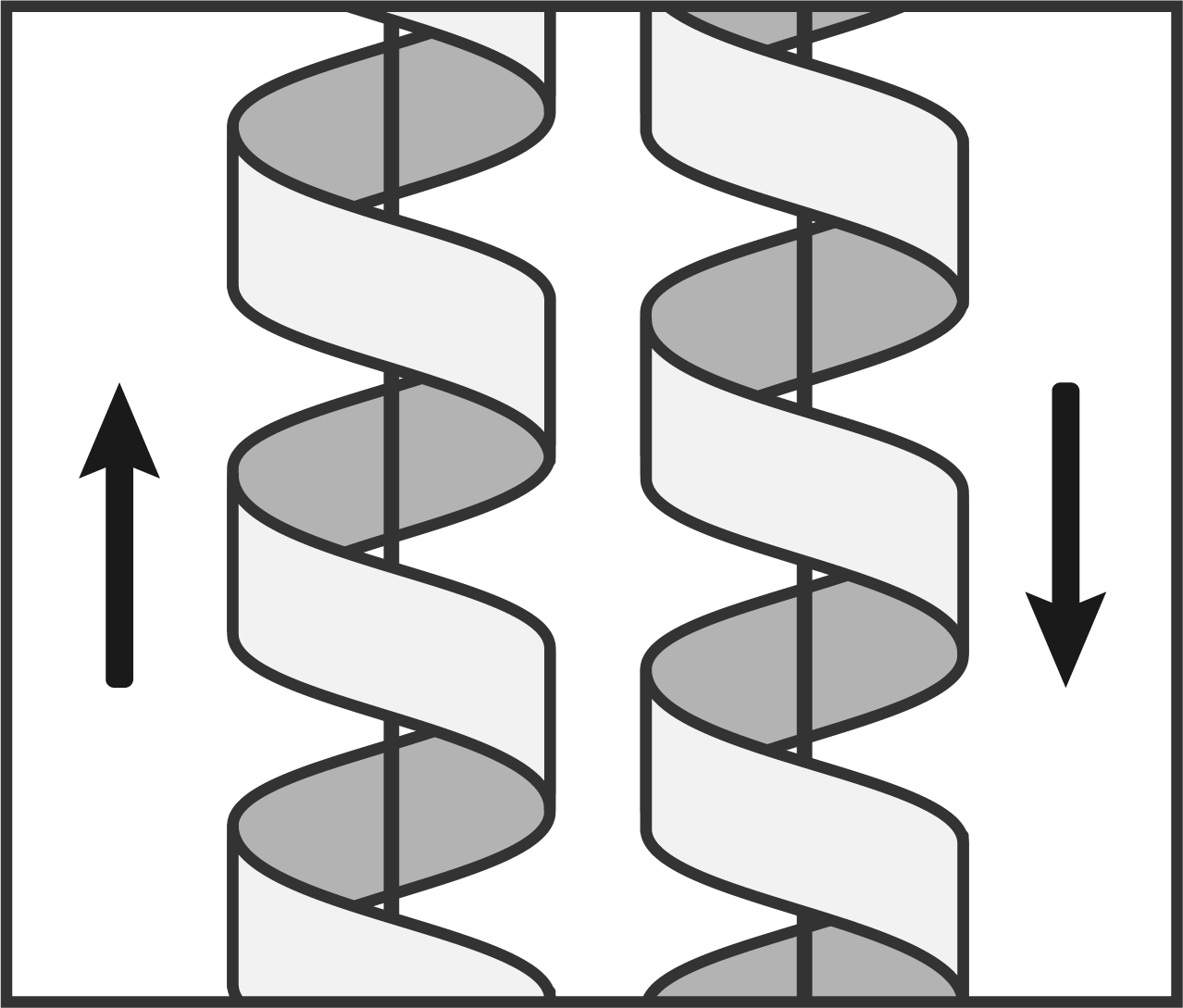}} \\
        OP &
        Oppositely handed,\newline
        aligned in parallel
        &
        \makebox[2cm][c]{\includegraphics[width=1.5cm]{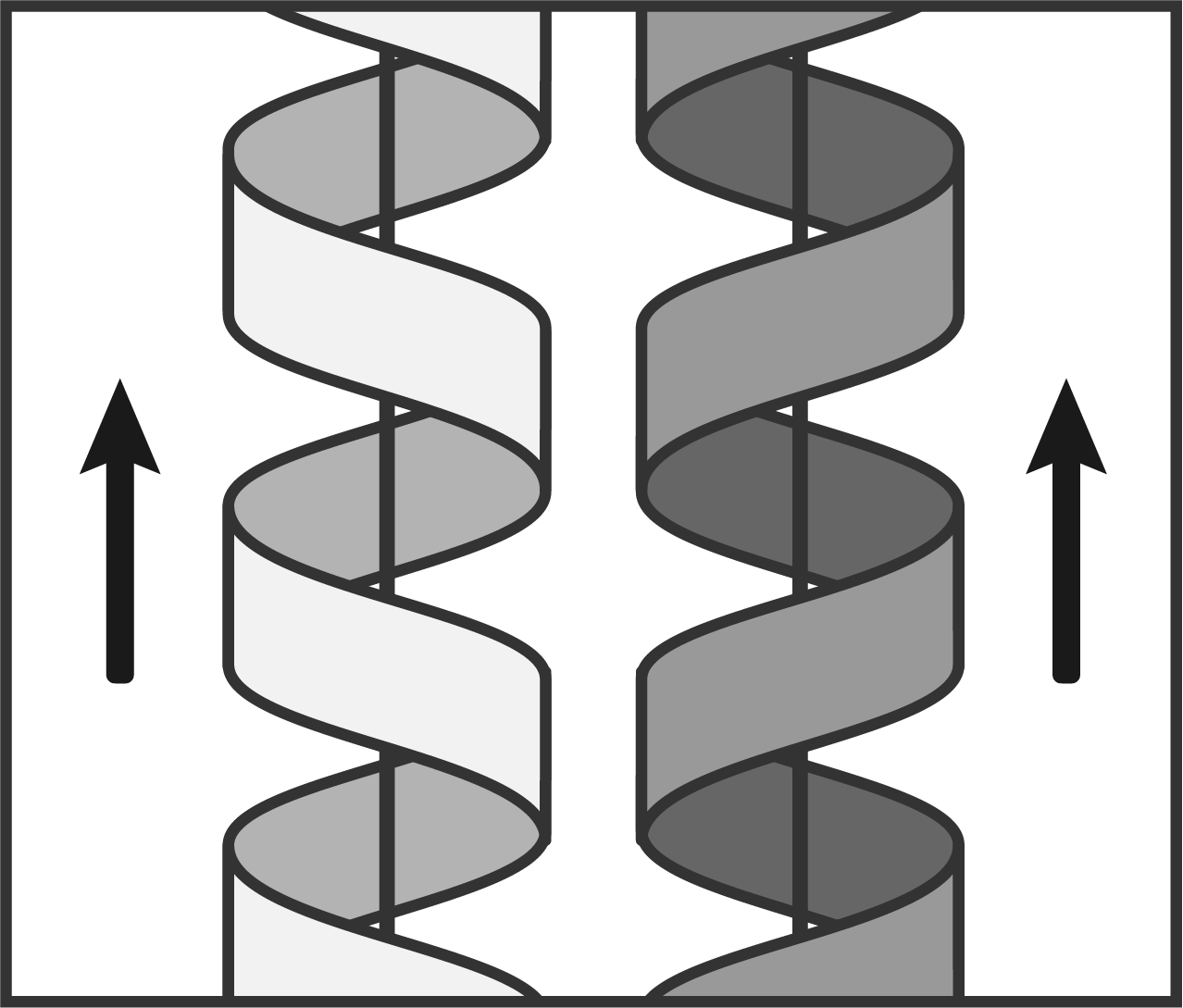}} \\
        OA &
        Oppositely handed,\newline
        aligned anti-parallel
        &
        \makebox[2cm][c]{\includegraphics[width=1.5cm]{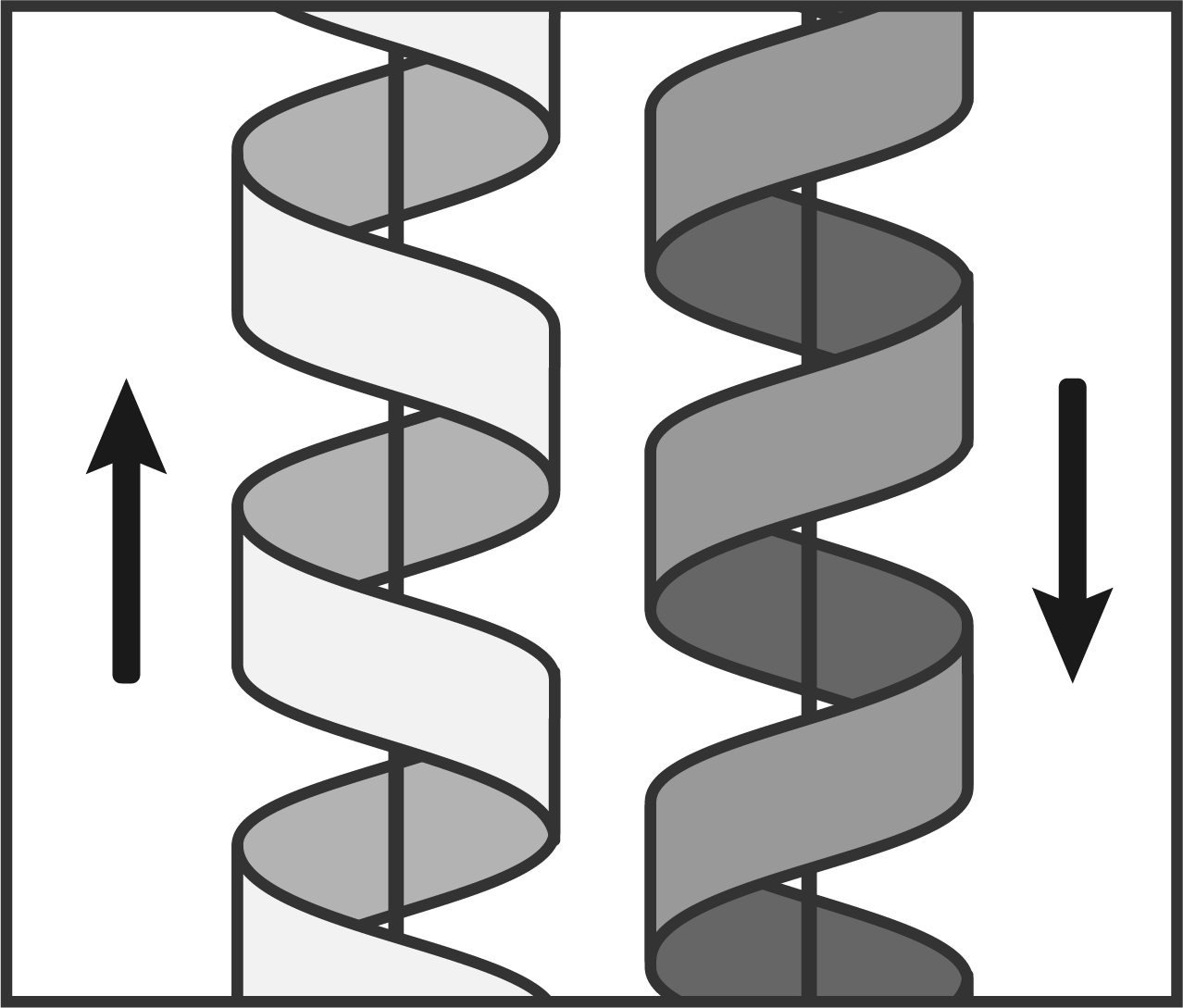}} \\
        \midrule
        \multicolumn{3}{l}{\textit{Multiple helix structures}} \\
        \midrule
        \{L$\uparrow$\} &
        \multicolumn{2}{m{7.5cm}}{All helices right-handed, pointing up} \\
        \{L$\uparrow \downarrow$\} &
        \multicolumn{2}{m{7.5cm}}{Right-handed helices with mixed up/down} \\
        \{L$\uparrow$--D$\uparrow$\} &
        \multicolumn{2}{m{7.5cm}}{Racemic mixture, all helices pointing up} \\
        \{L$\uparrow \downarrow$--D$\uparrow \downarrow$\} &
        \multicolumn{2}{m{7.5cm}}{Racemic mixture with mixed up/down} \\
        \{L$\uparrow$--D$\downarrow$\} &
        \multicolumn{2}{m{7.5cm}}{Correlated racemic ensemble: L-up, D-down} \\
        \bottomrule
    \end{tabular}
\end{table}

Each of these four interaction types is specified by four continuous variables: the inter-helical distance $R$, the rotation angles of the helices $\varphi_1$ and $\varphi_2$, and the relative lateral offset $\zeta$. 
While the distance $R$ is intuitively given as the separation between the two helix axes, the choice of how to define the angular variables $\varphi_1$, $\varphi_2$ and the shift parameter $\zeta$, requires a more careful consideration.
Without loss of generality, the first helix is considered to be right-handed with the carbonyl group pointing in positive $z$ direction, which is referred to as up-orientation.
This choice defines the relative coordinate system, which specifies the signs of $\varphi_1$, $\varphi_2$ and $\zeta$.
Their specific value is defined defined based on the positions of one of the 18 symmetry-equivalent nitrogen atoms within the repeat units of each of the helices. 
For the first helix, we select the nitrogen atom closest to the inter-helical distance vector. 
This choice confines $\varphi_1$ to a narrow range around the axis of minimal separation, specifically $[-10^\circ, +10^\circ)$. 
For the second helix, the reference nitrogen atom is chosen such that its axial displacement relative to the selected nitrogen of the first helix is positive but does not exceed $L_E$.
 This condition allows $\varphi_2$ to vary freely over $0^\circ$ to $360^\circ$, while restricting the lateral shift $\zeta$ to the interval $0 \leq \zeta < L_E$, ensuring only unique configurations along the helical axis are considered.

Instead of  considering the absolute rotation angle $\varphi_2$ to specify a pair configuration, we introduce a more convenient parameter:
\[
\chi = \varphi_1 - \tilde{h}\,\varphi_2 ,
\]

where $\tilde{h} = +1$ if both helices have the equal handedness, and $\tilde{h} = -1$ if their handedness is opposite. 
For helices of  equal handedness, $\chi$ is the angular difference between their reference atoms; for opposite handedness, it is the angular sum. 
This definition arises from the inherent screw symmetry of the helices: if two helices of  equal handedness are simultaneously rotated by integer multiples of $\phi_E=100^\circ$, the resulting configuration is equivalent to the original one, differing only by the choice of the reference atom as now another nitrogen atom of the first helix is closest to the inter-helical distance vector. 
In the case of helices with opposite handedness, the same invariance holds if the helices are rotated by equal amounts in opposite directions.

Formally, the parameter $\chi$ may be chosen within any interval covering a full rotation of $360^\circ$, such as $[0^\circ,360^\circ)$ or $[-180^\circ,180^\circ)$. 
In our case, however, we restrict the range differently, making use of an additional symmetry consideration:
For two helices that are aligned in parallel  (not antiparallel), the pair interaction obeys not only the helical symmetry discussed above, but also an additional invariance that arises from the fact that the two helices are identical and interchangeable. 
In fact, the choice of which helix is labeled first and which second is arbitrary. 
Hence, there exists a transformation between the two choices which formally changes the first and the second helix. 
For parallel aligned helices (equal or opposite handedness), this transformation corresponds to a reflection through the point $(0^\circ,-\phi_E/2)$ in the $(\varphi_1,\chi)$ plane. 
To explicitly include this symmetry in our analysis, we select the range of $\chi$ such that the inflection point lies in the middle of the chosen interval. 
This is achieved by restricting $\chi$ to the interval $[-230^\circ,130^\circ]$, which spans a full rotation of $360^\circ$ but is centered on the symmetry point at $\frac{1}{2}\phi_{E} = -50^\circ$. 
In this way, equivalent configurations related to labeling the helices in reverse order appear symmetrically within the parameter space, which simplifies both visualization and interpretation of the results.

\subsection{Ensemble Simulation}
To study low-energy structures based on pair interactions between all involved helices, we considered ensembles of 160 helices in a square simulation box with periodic boundary conditions. 
Each micro state of the ensemble is then defined by the 2D position of the helix centers ($x_i$ and $y_i$), the rotation angles ($\phi_i$), and the vertical shifts ($z_i$) of all helices. 
All of these parameters are defined with respect to a global coordinate system.
Again, the rotation angle and height displacement are specified with respect to one of the 18 symmetry-equivalent nitrogen atoms.
In addition, each helix has a handedness $h_i$ (1 for right-handed, and $-1$ for left-handed) and a direction $d_i$ ($+1$ for up, $-1$ for down). 
Based on these parameters, the contribution from each pair interaction can be obtained by transforming into the proper reference system through translation, rotation and mirroring such that one helix is right-handed point up.

To assess the phases that can be formed from parallel and anti-parallel aligned L-PA and/or D-PA, we considered enantiopure ensembles as well as racemic mixtures, with purely up-aligned systems and mixed up-and-down alignments. 
This results in four fundamental ensembles to which we refer as \{L$\uparrow$\}, \{L$\uparrow\downarrow$\}, \{L$\uparrow$-D$\uparrow$\}, and \{L$\uparrow\downarrow$-D$\uparrow\downarrow$\}.  
In addition, we consider a specific case where all right-handed helices are aligned upwards while left-handed ones are aligned downwards, which we refer to as \{L$\uparrow$-D$\downarrow$\}.  Table~\ref{tab:nomenclature} summarizes the individual ensembles. 

Low-energy structures of the considered ensembles were obtained from random initial configurations using simulated annealing with the Metropolis algorithm. To explore the configuration space, we applied four ergodic transformations: in-plane displacement, vertical displacement, rotation, and swaps of helices with distinct chiralities or orientations. These moves were randomly selected in a 3:3:3:1 ratio to ensure efficient sampling. 
The simulations used a pseudo-exponential annealing schedule: starting at $T_{\rm start} = \SI{10000}K$, we performed $1.5 \times 10^6$ Metropolis steps at each temperature, followed by a temperature reduction by a factor of $\tau = 0.9$ until the simulation temperature reached a value below $T_{\rm end}=\SI{0.01}K$.

In this study, we focus only on the properties of the low-energy configurations obtained at the end of the simulation. 
The specific paths to these states are not physically meaningful, because relative diffusion rates are not reflected in the chosen ration between update types.
However, some insights can be gained from the simulation process, particularly from the specific heat, see Figure~\ref{fig:HeatCap}, whose features can be attributed to the formation of structural properties observable in the final films.
For completeness, representative results from the simulated annealing simulations are provided in the Supplementary Information, Section \ref{sec:Waermekapazitaet}. 

\section{Results and Discussion}
\subsection{Pair-Interaction Potentials}
\subsubsection{Distance-Dependent Interaction Energy at Frozen Relative Orientation}

As defined in Section 2, the interaction energy between two helices depends on four continuous geometric parameters: the interaxial distance $R$, the azimuthal angle $\varphi_1$ of the first helix, the relative angle $\chi$ (defined as the sum or difference of the two azimuthal angles, depending on handedness), and the relative vertical offset $\zeta$. 
Additionally, different combinations of handedness and axial direction result in the four distinct interactions: EP, EA, OP, and OA. 
The four-dimensional parameter space $(R,\varphi_1,\chi,\zeta)$ makes direct visualization of the interaction energy challenging. 
To reduce complexity, we begin by examining the dependence on $R$ while freezing the other degrees of freedom, thereby isolating distance regimes where relative orientation significantly influences the interaction.

At large separations, local contacts between functional groups---whose spatial arrangement is determined by relative orientation---play a minor role, and the interaction energy approaches zero. 
In contrast, at short to intermediate distances, steric complementarity and the possibility of interdigitation dominate the interaction. 
To explore these effects, we computed distance-dependent binding energy profiles for a comprehensive set of fixed relative orientations. 
Specifically, for each combination of $\varphi_1$, $\chi$, and $\zeta$ ($10 \times 18 \times 10 = 1800$ configurations), we evaluated $E_{\text{bind}}(R)$ for $R$ ranging from sterically forbidden overlaps up to \SI{20}\angstrom, beyond which interactions are negligible. 
The binding energy $E_{\text{bind}}(R)$ is obtained as the difference between the total energy of the dimer and twice the total energy of an isolated helix.

\begin{figure}[!htbp]
\centering
\includegraphics[width=0.9\textwidth]{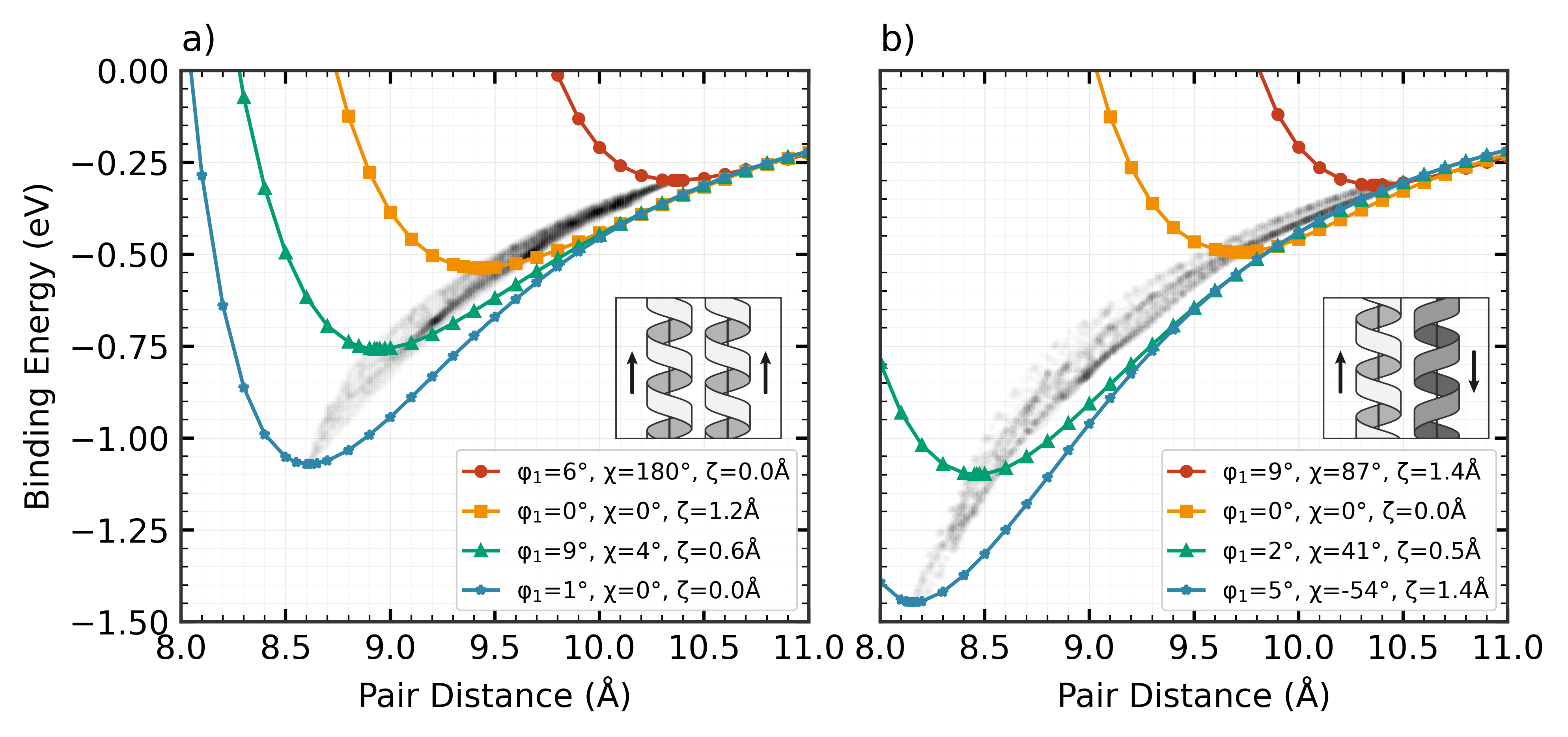}
\caption{Distance-dependent binding energy profiles for representative configurations under fixed orientation parameters ($\varphi_1, \chi, \zeta$). (a) Four cases from the OA class; (b) Four cases from the EP class. Gray curves in the background correspond to all other sampled configurations, illustrating the overall variability across the orientation space. The legends indicate ($\varphi_1$, $\chi$, $\zeta$) values for each highlighted curve. These results demonstrate that opposite-handedness (OA) enables stronger binding and closer contact than same-handedness (EP) arrangements.}
\label{fig:Ropt_EP_OA}
\end{figure}

Figure~\ref{fig:Ropt_EP_OA}a shows binding energy $E_{\text{bind}}(R)$ curves for four selected relative orientations in EP alignment (colored lines), with the corresponding ($\varphi_1$, $\chi$, $\zeta$) values indicated in the legend. 
These selections illustrate the diversity in both optimal binding distance $R_{\text{opt}}$ and minimum energy $E_{\text{min}}$ arising from differences in relative orientation. 
The $E_{\text{bind}}(R)$ profiles are overlaid on the complete set of equilibrium distances $R_{\text{opt}}$ and minimum energies $E_{\text{min}}$ for all 1800 relative orientations (gray dots). Three key observations emerge:(i) For all configurations, the energy curves exhibit a well-defined minimum, with equilibrium spacing between \SI{8.6}{\angstrom} and \SI{10.3}{\angstrom}; (ii) The depth of the corresponding energy minima varies from \SI{-1.07}{eV} to \SI{-0.25}{eV}; (iii) At larger distances ($R > \SI{10.5}{\angstrom}$), the curves converge and orientation-dependent differences become negligible, confirming that angular effects dominate primarily in the near-contact regime.

Figure~\ref{fig:Ropt_EP_OA}b presents corresponding results for OA alignment, where helices possess opposite handedness and are oriented antiparallel. 
While the overall shape of the energy curves remains similar to EP alignment, two notable differences emerge. 
First, the minimal equilibrium spacing shifts toward smaller values, with $R_{\text{opt}}$ reaching as low as \SI{8.16}{\angstrom}. 
Second, the interaction strength is significantly enhanced: the most favorable configurations exhibit binding energies up to \SI{-1.45}{eV}, approximately \SI{400}{meV} deeper than the strongest EP configurations. 
These findings indicate that opposite-handed arrangements enable closer approach and stronger stabilization, consistent with Wallach's rule for isolated $\alpha$PA strands.

The remaining interaction possibilities, EA and OP, are presented in the SI, where similar qualitative trends are observed (Figure~\ref{fig:Ropt_EA_OP}). 
For EA alignment, minimal distances for frozen orientations range from \SI{8.5}{\angstrom} to \SI{10.3}{\angstrom}, with corresponding binding energies between \SI{-1.04}{eV} and \SI{-0.25}{eV}, whereas OP configurations exhibit slightly smaller equilibrium distances up to \SI{8.3}{\angstrom}, accompanied by stronger binding energies reaching up to \SI{-1.30}{eV}. These results further support that relative handedness and orientation govern both equilibrium spacing and interaction strength between helices.
For better comparability of the energy ranges and obtained equilibrium distances under frozen azimuthal angles and offsets, Figure~\ref{fig:Ropt_all} overlays the boundary of the obtained distributions.

For interaxial distances larger than \SI{11}{\angstrom}, the interaction energy approaches zero, independent of the specific frozen configuration within a given alignment and is nearly identical across all four classes (EP, OA, EA, and OP). Based on this observation, the long-range portion of the computed energy curves was fitted to a $1/R^6$ term and subsequently extrapolated to account for distances beyond the sampled range.
\subsubsection{Dependence on Relative Offset}

We now analyze the interaction as a function of the relative axial displacement between two helices. 
For each combination of $\varphi_1$ and $\chi$, the relative vertical offset $\zeta$ was varied from 0 to $L$, while the interaxial distance $R$ was always chosen to correspond to the minimal value for the considered set of $\varphi_1$, $\chi$, and $\zeta$. 
This approach allows visualization of the effect of steric repulsion of the functional methyl groups, which can either permit or prevent interdigitation.

\begin{figure}[!htbp]
\centering
\includegraphics[width=0.9\textwidth]{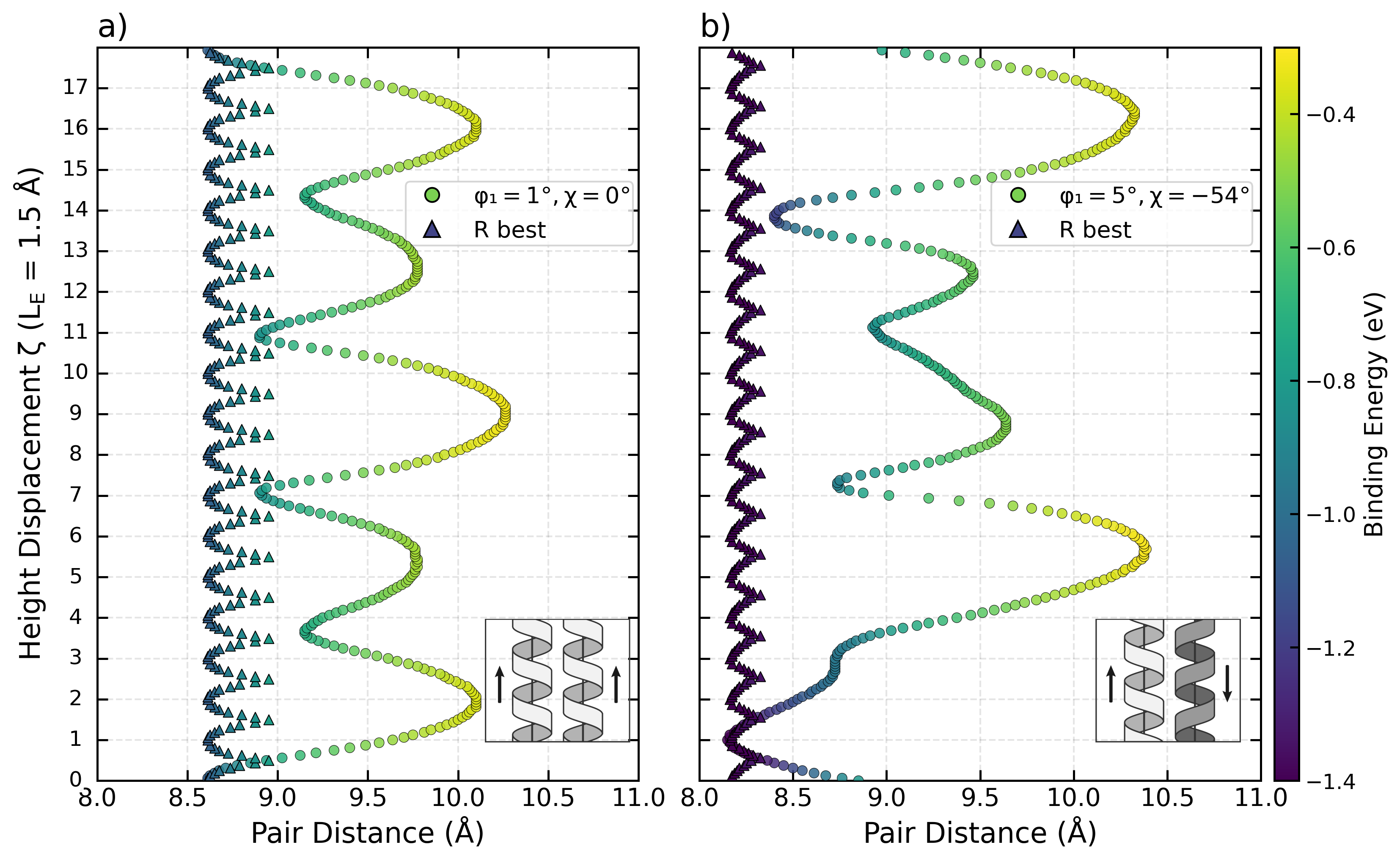}
\caption{Optimized pair distance $R$ for fixed $\varphi_1$ and $\chi$ as a function of the relative vertical offset $\zeta$ for EP (left) and OA (right) alignment. Each colored circle represents the binding energy for a given $\varphi_1$-$\chi$ combination, with the color scale indicating the magnitude of the interaction energy (blue: stronger binding, red: weaker binding). Triangles indicate the minimal pair distance $R_{\text{best}}$ for each $\zeta$, corresponding to the most favorable angles $\varphi_1$ and $\chi$. The oscillatory dependence on $\zeta$ highlights the critical role of axial registry in achieving optimal interdigitation and binding strength.}
\label{fig:zeta_EP_OA}
\end{figure}

Results for EP and OA configurations with $\varphi_1 = \SI{0}{\degree}$ and $\chi = \SI{0}{\degree}$ are shown in Figure~\ref{fig:zeta_EP_OA} as circles, where the horizontal axis corresponds to the optimal pair distance and the vertical axis to offset $\zeta$. 
The color code indicates the binding energy for each configuration. 
Additionally, the graphs feature the distance and energy for each offset $\zeta$ where $\varphi_1$ and $\chi$ yield minimal binding energy (triangles). 
Due to the helical symmetry of each helix, a shift of $L_E$ is equivalent to a rotation of \ang{100}. 
Therefore, the curves presented in Figure~\ref{fig:zeta_EP_OA} represent configurations with $\varphi_1$ being multiples of \SI{20}{\degree} but displaced by the corresponding lateral shift.

For EP alignment (Figure \ref{fig:zeta_EP_OA}a), the closest configuration is obtained for zero relative offset and vanishing angles $\varphi_1$ and $\chi$. 
This result can be explained by the fact that for equally oriented, parallel helices of the same handedness, the helical structure adopts an interdigitated configuration. 
Upon a lateral shift of half the pitch, interdigitation is lost due to steric repulsion between helix backbones. 
This repulsion is strongest when methyl groups face each other, occurring in configurations such as ($\varphi_1 = \SI{0}{\degree}$, $\chi = \SI{180}{\degree}$, $\zeta = \SI{0.0}{\angstrom}$) or equivalently ($\varphi_1 = \SI{0}{\degree}$, $\chi = \SI{0}{\degree}$, $\zeta = 9 L_E$) due to symmetry. As shown in Figure~\ref{fig:zeta_EP_OA} (left), this configuration corresponds to the largest $R_{\text{opt}}$. 
Moreover, due to additional symmetry for parallel-aligned helices, the graphs exhibits mirror symmetries with respect to interdigitate ($\zeta=0$) and non-interdigitate configurations ($\zeta=9 L_E$).

For the OA system (Figure~\ref{fig:zeta_EP_OA}b), similar behavior is observed. 
Upon shifting the two helices for fixed angles, the optimal distance and corresponding binding energy oscillate between interdigitated configurations with smaller optimal distances and lower binding energies, and non-interdigitated configurations with larger distances up to \SI{10.5}{\angstrom}.
In contrast to the EP case, the curve lacks  symmetry due to the missing symmetry of anti-parallel aligned helices.
Furthermore, configurations of smallest and largest $R_{\text{opt}}$ are not related via a \SI{180}{\degree} rotation of one helix, as different helix orientations result in different functional group arrangements.

Similar observations for EA and OP cases, including symmetry properties for parallel alignment and closer distances for opposite-handed pairs, are presented and discussed in the Supplementary Information (Figure~\ref{fig:zeta_EA_OP}). 
In particular, the additional symmetry with respect to a shift of \SI0{\angstrom} and $9L_E$ for parallel alignment which is missing for antiparallel alignment is evident.  

\subsubsection{Angle Dependence}

Having analyzed the interaction potential of helix pairs with respect to distance and offset at fixed angles, we now consider the energy landscape regarding angular orientation. 
As presented in Section 2, symmetry properties of the configuration space allow $\varphi_1$ to be limited to \SI{-10}{\degree} to \SI{10}{\degree} while $\chi$ spans a full rotation of \SI{360}{\degree}. To incorporate the additional symmetry point for parallel-aligned helices at ($-\varphi_E/2$), we choose the relative angle parameter $\chi$ (angle difference for same-handedness helices and angle sum for opposite-handedness helices) in the range \SI{-230}{\degree} to \SI{130}{\degree}.

\begin{figure}[!htbp]
\centering
\includegraphics[width=1\textwidth]{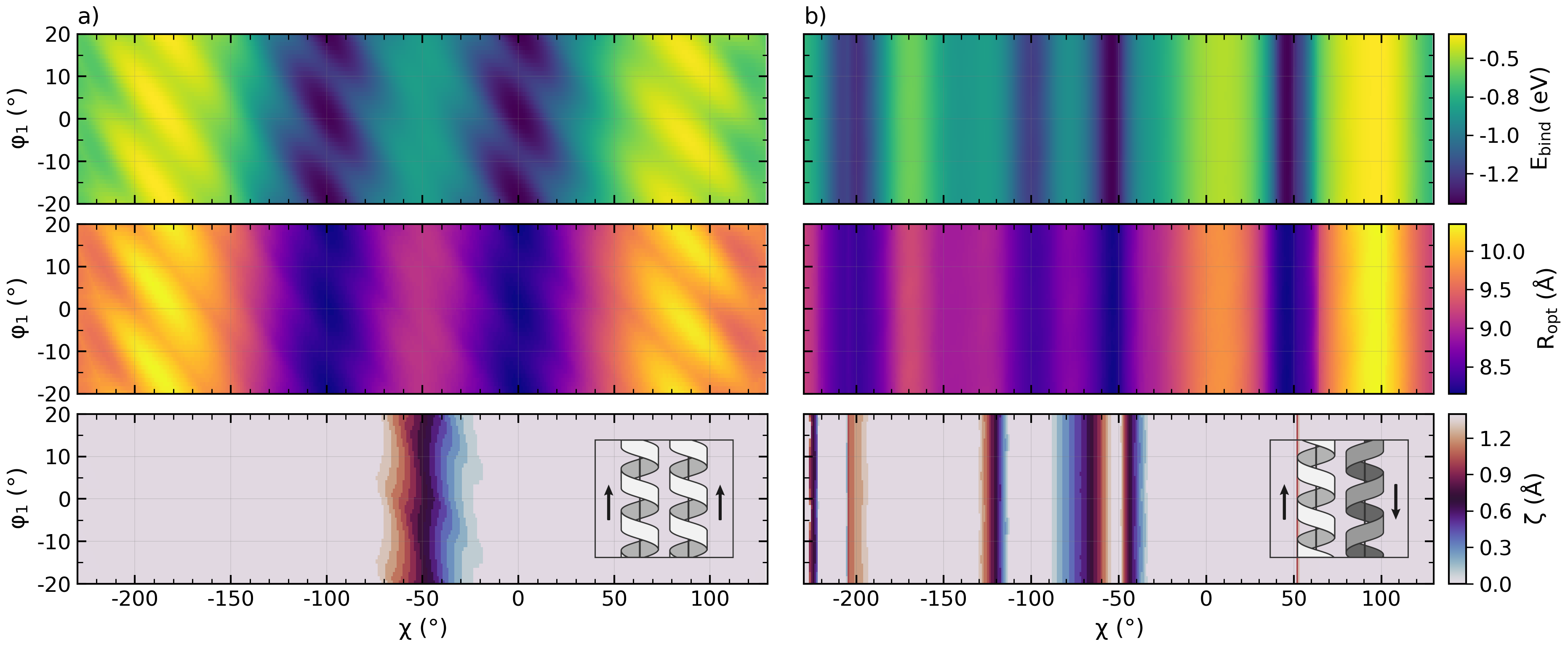}
\caption{Heatmaps of binding energy (top), equilibrium distance $R_{\text{opt}}$ (middle), and relative offset $\zeta_{\text{opt}}$ (bottom) as a function of angular parameters $\varphi_1$ and $\chi$ for EP (left) and OA (right) configurations. The EP configuration shows complex 2D dependence on both angles, while the OA configuration depends primarily on the relative sum $\chi$, revealing a fundamental difference in how handedness dictates the interaction landscape.}
\label{fig:heatmap_EP_OA}
\end{figure}

Figure \ref{fig:heatmap_EP_OA} shows the obtained binding energies with corresponding values of $R_{\text{opt}}$ and $\zeta_{\text{opt}}$ in heat map format.
For the EP configuration (Figure~\ref{fig:heatmap_EP_OA}, left), the heatmaps reveal how interaction energy, equilibrium distance, and relative offset depend jointly on angular variables $\varphi_1$ and $\chi$. 
The binding energy map (top) exhibits pronounced two-dimensional features, indicating both angular parameters substantially shape the interaction. 
Smooth variation across both $\varphi_1$ and $\chi$ with binding energy variations up to \SI{0.6}{\electronvolt} demonstrates that in EP alignment, orientation cannot be reduced to a single dominant parameter but emerges from genuinely two-dimensional angular dependence.

Symmetry-related structures can be identified at inflection points around $\varphi_1 \approx \SI{-10}{\degree},\SI{0}{\degree}$ combined with $\chi = \SI{-50}{\degree},\SI{130}{\degree}$, and \SI{-230}{\degree}.
These equivalent points arise from intrinsic helical symmetries and involve compensating shifts in axial offset by $L_E$, as reflected in the offset map (bottom).
 A distinct global minimum is observed near $\varphi_1 \approx \SI{1}{\degree}$, $\chi \approx \SI{1}{\degree}$ (and correspondingly near $\varphi_1 \approx \SI{-1}{\degree}$, $\chi \approx \SI{-99}{\degree}$), with optimal distance $R_{\text{opt}} \approx \SI{8.7}{\angstrom}$ and zero offset ($\zeta_{\text{opt}}=0$).

In contrast, the OA configuration (Figure~\ref{fig:heatmap_EP_OA}, right) shows markedly different features. Here, binding energy maps are dominated by stripe-like features aligned along $\varphi_1$, demonstrating interaction insensitivity to the absolute value of $\varphi_1$. 
Instead, the dominant dependence is on the relative sum $\chi=\varphi_1+\varphi_2$, consistent with opposite helicity. 
Due to anti-parallel alignment, inversion symmetry with respect to the configuration $(\varphi=\SI{0}{\degree}, \chi=-\phi_E=\SI{-50}{\degree})$ present in the EP case is lost. 
Instead, two preferential orientations with binding energies of about \SI{-1.44}{\electronvolt} stand out at $\chi \approx \SI{46}{\degree}$ and $\chi \approx \SI{-54}{\degree}$, with additional local minima around $\chi \approx \SI{-100}{\degree}$ and \SI{-200}{\degree}. 
Equilibrium distances and offsets reflect these patterns, with sharp transitions in $\zeta_{\text{opt}}$ aligning with changes in favorable angular configurations.

Corresponding analyses of EA and OP interactions are shown and discussed in the Supplementary Information, where qualitatively similar observations are found: well-defined global minimum configurations for EA alignment Figure~\ref{fig:heatmap_EA_OP}a and stripe-like features aligned along $\varphi_1$  due to the opposite-handedness and inversion symmetry in the ($\varphi_1,\chi$)-plane with respect to (\SI{0}{\degree}$,\SI{-50}{\degree}$) because of the parallel alligment for the OP system (Figure~\ref{fig:heatmap_EA_OP}b)

In summary, the pair interaction analysis reveals that OA and OP configurations, involving helices of opposite handedness, consistently allow closer approach (smaller $R_{\text{opt}}$) and stronger binding (more negative $E_{\text{min}}$) than EP or EA configurations. This fundamental difference in pairwise stability, dictated by relative handedness and orientation, is the key factor governing structural properties of larger self-assembled films discussed in the following section.


\subsection{Global Minimum Configurations of Helix Pairs}

After analyzing the dependence of the interaction potential on distance, offset, and angular orientation, we now aim at identifying which specific interactions stabilize the most favorable helix arrangements. 

\begin{figure}[!htbp]
\centering
\includegraphics[width=0.99\textwidth]{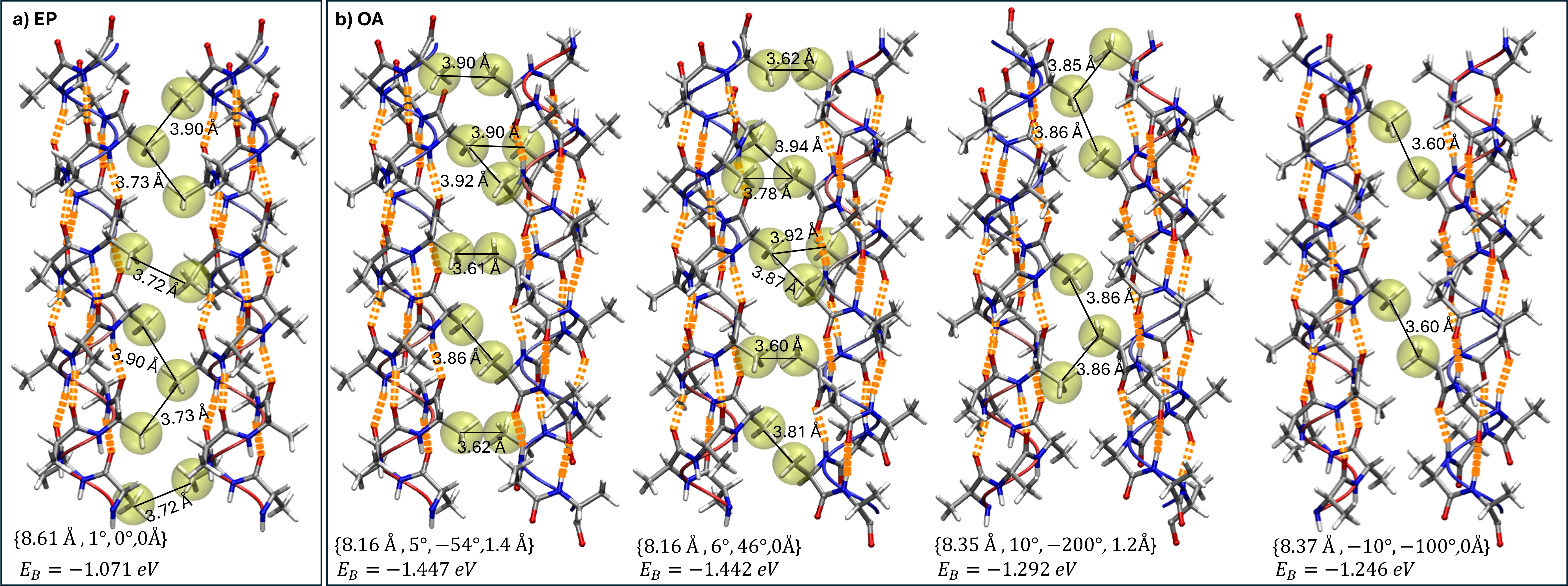}
\caption{Local Minimum configurations of EP (a) and OA (b) helix pairs, where  for better recognition the helical backbone is represented as a coil. The specific relative configuration parameters {$R$,$\varphi_1$,$\chi$,$\zeta$} as well as the corresponding binding energies $E_B$ are given below each configuration. Intra-helical hydrogen bonds stabilizing the $\alpha$ helix are indicated in orange. Methyl groups with inter-helical distances smaller than $\SI{4.0}{\angstrom}$ are highlighted by yellow spheres with the individual distances given in the figure.}
\label{fig:VMD}
\end{figure}

Figure \ref{fig:VMD} displays selected configurations of EP (a) and OA (b) helix pairs, which correspond to local minima of the interaction potentials presented in Section 3.1. 
The representations show the complete unit cell comprising 18 alanine residues over five helical turns of each helix. 
For better visualization of the individual handedness and up/down orientation, the helical backbone is indicated by a coil.
Furthermore, the intra-helical hydrogen bonds stabilizing the $\alpha$-helical structure are shown as orange dashed lines. 
To characterize the intramolecular interaction, the functional methyl groups with  distances to groups of the other helix of less than $\SI{4.0}{\angstrom}$ are highlighted by transparent yellow spheres. 
The corresponding plots for  of EA and OP are presented in SI (Figure~\ref{fig:SI_VMD}).

For the EP system, the global minimum corresponds to a configuration with $\varphi_1 = \SI{1}{\degree}$, $\chi = \SI{0}{\degree}$, and $\zeta = \SI{0.0}{\angstrom}$ which can be interpreted as if one helix is only horizontally displaced from the other.
This consideration yields an explanation why it constitutes the energetically most favorable configuration as it allows the best way to interdigitate the helices allowing for dense packing. 
Any relative rotation or offset of one of the helices would lead to unfavorable interactions between functional groups, increasing the  helix-helix distance and thus reducing the overall binding strength.
As shown in Figure~\ref{fig:VMD}a),  in this  confirmation there are six methyl-methyl contacts with distances ranging from $\SI{3.7}{\angstrom}$ to $\SI{3.9}{\angstrom}$. 
In line with previous quantum-chemical analyses of methyl-methyl contacts published by Dutta et al~\cite{Dutta2025-fy},  these intermolecular  distances are consistent with the presence of weak but attractive dispersion-driven interactions with energies of approximately $-0.02$ to $\SI{-0.07}{eV}$ per contact ($-2$ to $\SI{-7}{kJ mol^{-1}}$ ).

For the global minimum of the OA interaction (Figure~\ref{fig:VMD}b, left), there are also six methyl-methyl contacts with distances smaller than $\SI{4.0}{\angstrom}$. 
Here, however, these distances are smaller than for EP and approach the ideal distance reported by Dutta et al.~\cite{Dutta2025-fy}. 
For the second most stable local minimum configuration (Figure~\ref{fig:VMD}b, middle-left) similar interactions are observed, resulting in a binding energy nearly identical to the global minimum.
The remaining two local minimum configurations of the OA alignment (Figure~\ref{fig:VMD}b, middle-right and right), on the other hand, exhibit only four and two interaction points and thus a lower number of interhelical methy-methyl distances smaller than $\SI{4.0}{\angstrom}$

Comparing the structural properties of these minimum configurations and their related individual energies, the following conclusions are drawn: 
On the one hand, the local interaction of the closest motifs are between the non-polar methyl groups. 
The distances of polar groups of different helices, such as the amide and the carbonyl group, are too large to form hydrogen bridges. 
Hence, the stabilizing interaction is of van-der-Waals type rather than  hydrogen bonding as previously assumed~\cite{ha2020}. 
As van-der-Waals interactions do not alter the electronic density of each helix, it can be concluded, that the attractive interactions between helices do not disturb the electronic structure of each individual helix.
This finding provides a justification, that for studies focusing on the electronic transport, the consideration of isolated single helices is sufficient. 
In particular,  local dipole moments concluded from models of isolated helices are valid to describe the properties of helix bundles or films.
On the other hand,  the total binding energy is $\SI{-1.0}{eV}$ up to $\SI{-1.4}{eV}$, i.e. up to 20 times larger than the optimal interaction between  two non-bonded methyl groups.
As only six or less of such interactions are present,  it is concluded, that the stabilization of the individual configuration is not due to a few specific local interactions, but rather of the systems as a whole.
This conclusion is strengthened by the fact that local minima of OA with only two interacting methyl group pairs (Figure~\ref{fig:VMD}b) right) shows stronger binding  than the  EP configuration with six of such interactions.



\subsection{Low-Energy Configurations of Self-Assembled Films}

After analyzing the interaction properties of isolated helix pairs, we now examine ensembles to investigate how these pairwise interactions manifest in larger assemblies.
This allows us to assess whether optimal pair configurations are preserved in collective structures or whether frustration effects and packing constraints drive the system into alternative arrangements.

\begin{figure}[!htbp]
\centering
\includegraphics[width=1\textwidth]{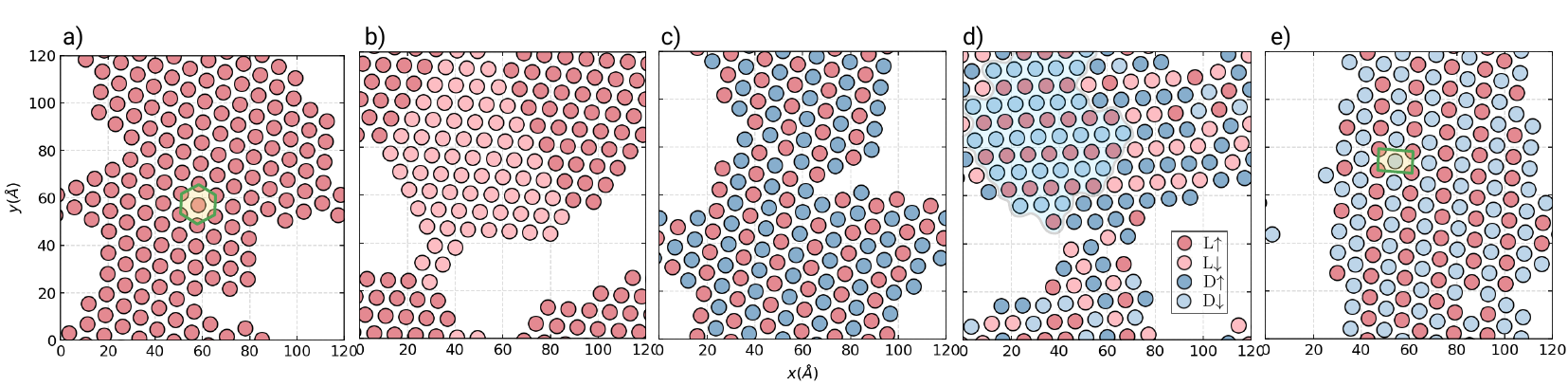}
\caption{Representative configurations of self-assembled films for the different systems defined in Table~\ref{tab:nomenclature}: (a) \{L$\uparrow$\}, (b) \{L$\uparrow \downarrow$\}, (c) \{L$\uparrow$-D$\uparrow$\}, (d) \{L$\uparrow \downarrow$-D$\uparrow \downarrow$\}, and (e) \{L$\uparrow$-D$\downarrow$\}.}
\label{fig:configs_all}
\end{figure}

To investigate properties of low-energy arrangements that  reflect features of experimentally observed self-assembled structures, we analyze structural aspects of configurations obtained from heuristic optimization using simulated annealing. 
Figure~\ref{fig:configs_all} shows representative 2D structures for each of the five ensembles considered, with notable features that will be discussed in detail.

To complement the structural characterization, we performed statistical analysis over the best film configurations obtained from 100 independent runs for each system. 
We focus on structural properties, particularly the radial distribution function (RDF) and the relative orientation of nearest neighbors (RONN), by averaging over all 100 configurations. 
Considering that each ensemble consists of 160 helices in close packing with up to 6 nearest neighbors each, the statistical analysis includes over \num{30000} individual pair arrangements, providing a solid basis for characterizing dominant film features.

\subsubsection{Structural Properties of Parallel Enantiopure Films}

\begin{figure}[!htbp]
\centering
\includegraphics[width=1\textwidth]{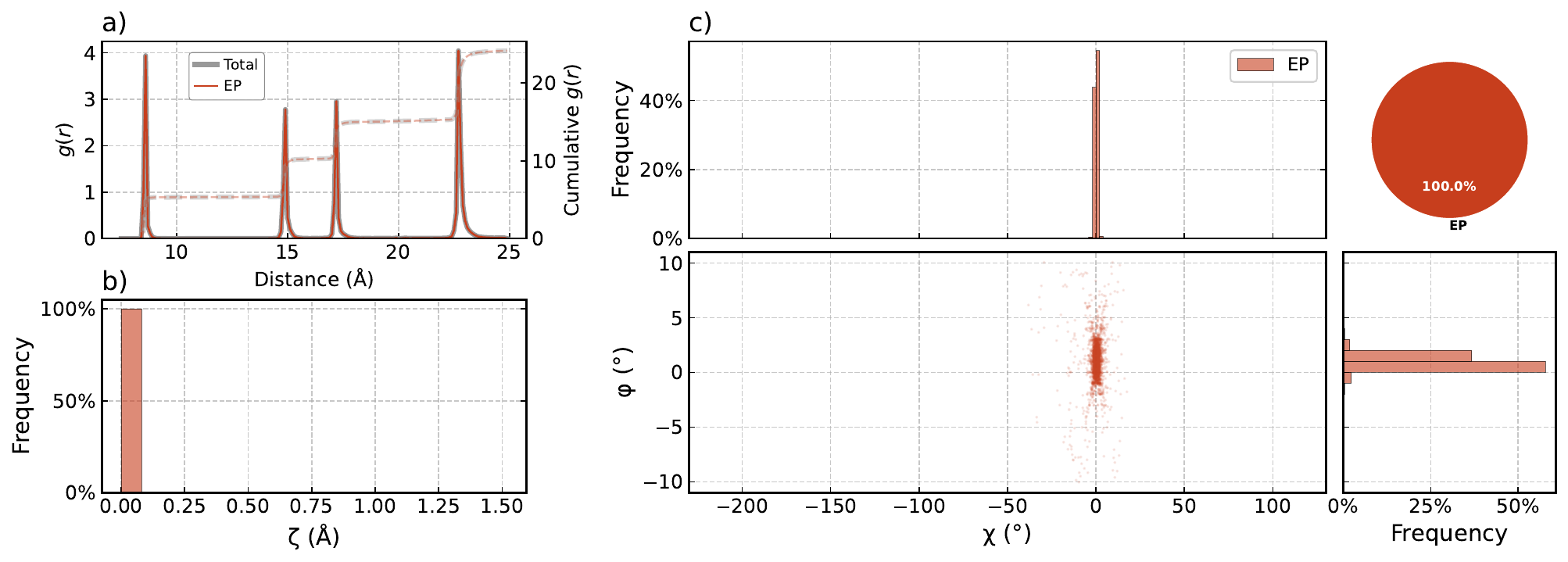}
\caption{Structural analysis of the \{L$\uparrow$\} system. (a) Radial distribution function (RDF) showing perfect hexagonal packing. (b) Distribution of the angle $\varphi_1$ for nearest neighbors. (c) Joint distribution of the angle difference $\chi_S$ and relative axial offset $\zeta$ for nearest neighbors. The data confirms a frustration-free, homogeneous structure where every pair interaction is in its optimal EP configuration.}
\label{fig:RONN_RDF_EP}
\end{figure}

For the L$\uparrow$ system (all helices of equal handedness and parallel alignment), all pair interactions are of type EP. 
As shown in Figure \ref{fig:configs_all}a, simulations result in a hexagonal pattern. 
Analysis of the RDF shown in Figure \ref{fig:RONN_RDF_EP}a confirms perfect hexagonal alignment. 
Well-defined peaks occur at \SI{8.6}{\angstrom}, \SI{14.9}{\angstrom}, \SI{17.2}{\angstrom}, and additional positions. These ratios reflect characteristics of an ideal hexagonal lattice. 
Based on the nearest-neighbor distance $a_0 = \SI{8.6}{\angstrom}$, the expected second- and third-nearest neighbor distances are precisely at $\sqrt{3}a_0 = \SI{14.9}{\angstrom}$ and $2a_0 = \SI{17.2}{\angstrom}$, respectively. 
Comparing this finding with the pair interaction potential reveals that this distance of approximately \SI{8.6}{\angstrom} represents the lower boundary of the optimized distance between two isolated helices in EP configuration (see Section 3.1), indicating very dense packing.

To characterize RONN, $\varphi_1$, $\chi$, and $\zeta$ of all pairs with distances smaller than \SI{10}{\angstrom} were statistically examined. 
Figure \ref{fig:RONN_RDF_EP}b shows the distribution of offset $\zeta$, which is sharply peaked at \SI{0}{\angstrom}.
Figure \ref{fig:RONN_RDF_EP}c presents the joint distribution of angle $\varphi_1$ and relative angle $\chi$ as a scatter plot with projected histograms along the corresponding axes. 
The results demonstrate that essentially all nearest-neighbor contacts occur at $\chi \approx \SI{0}{\degree}$ and $\zeta \approx 0$, indicating strict angular arrangement.
This observation underlines the high degree of structural homogeneity in this ensemble and shows that dense hexagonal packing is realized exclusively through energetically optimal EP contacts.
This frustration free arrangement is only possible due to the fact that the helical symmetry of $\alpha$PA reduces the value of $\varphi_1$ to a range with repetition every $20^\circ$. 
 Since in a perfect hexagonal lattice the angle between lattice vectors equals \SI{60}{\degree} (a multiple of \SI{20}{\degree}), a helix can adopt the optimal configuration with all its neighbors simultaneously.

To investigate the influence of up/down alignment on self-assembled structures, simulations of same-handed helices either oriented up or down were conducted (\{L$\uparrow \downarrow$\} system). 
From Figure~\ref{fig:configs_all}b, one observes: 
(i) a densely packed overall hexagonal arrangement; 
(ii) demixing of differently oriented helices into individual domains; and 
(iii) straight domain boundaries. 
Observation (ii) can be explained as follows: 
Since a pure EP system forms frustration-less films, where each nearest-neighbor interaction is in its global minimum, and any EA orientation is energetically less favorable than EP alignment with $\zeta = 0$ and $\chi = 0$ (see Section 3.1), the bonding energy of a helix is lowest when surrounded by six helices of the same alignment. 
For mixtures of up and down orientated helices, this causes the demixing. 
Observation (iii) can be understood by considering that at domain boundaries, each helix is in EA interaction with some of its  neighbors. 
For straight boundaries, each helix is surrounded by four helices of the same direction (EP interaction) and two of opposite alignment (EA interaction). 
Non-straight boundaries would require "corner points" with three neighbors of each type, which is energetically less favorable, leading to kink-free domain boundaries. 
Regarding observation (i), the cross-domain hexagonal structure can be explained by the fact that the most stable EA orientation occurs at \SI{8.65}{\angstrom}, similar to the EP lattice constant (see Section 3.1). 
That these EA arrangements are also in the best possible pair-interaction configuration is confirmed by RONN statistics in Figure~\ref{fig:RONN_RDF_EA}, showing a narrow distribution at $\chi=-\SI{77}{\degree} $ and $\zeta=\SI{0.3}\angstrom$ corresponding to the global-minimum arrangement of EA pairs. 
Due to boundary minimization, less than 6\% of all nearest neighbor interactions are of EA nature.

We conclude that enantiopure films tend toward structures of parallel alignment with identical angular orientation and vanishing vertical displacement, resulting in hcp structures. 
This confirms experimental observations reported for enantiopure L-PA films~\cite{ha2020}. 
It is particularly worth noting that, in contrast to our theoretical system of infinite chains, experimental $\alpha$PA is terminated by amine or carboxyl groups. 
Due to the chemical differences of these capping groups, constant-current STM measurements would result in apparent height differences. 
We therefore conclude that STM measurements also indicate parallel aliments in the enantiopure domains .

\subsubsection{Structural Properties of Films Formed from Racemic Mixtures}

\begin{figure}[!htbp]
\centering
\includegraphics[width=1\textwidth]{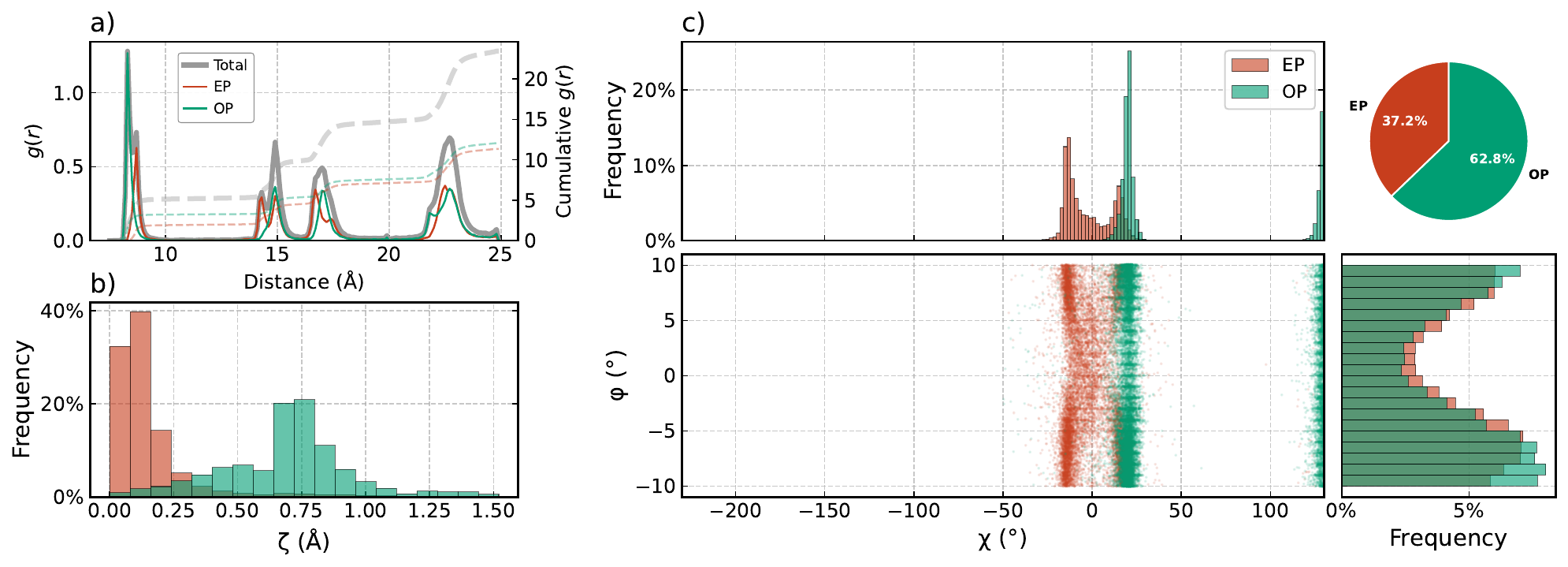}
\caption{Structural analysis of the \{L$\uparrow$-D$\uparrow$\} system. (a) Radial distribution function (RDF) showing the two-peak feature indicating different EP and OP neighbor distances. (b) Distribution of the relative offset $\zeta$ for EP interactions. (c) Joint distribution of the angle difference $\chi$ and relative axial offset $\zeta$ for EP interactions. The loss of a well-defined optimal orientation for EP pairs indicates frustration induced by the preferred, incompatible OP configuration.}
\label{fig:RONN_RDF_OP}
\end{figure}

For the \{L$\uparrow$-D$\uparrow$\} system (racemic mixture, all helices up-aligned), Figure~\ref{fig:configs_all}c indicates an overall hexagonal structure where right- and left-handed helices are arranged in a line pattern.
 Analysis of the RDF (Figure~\ref{fig:RONN_RDF_OP}a) reveals that, in contrast to the enantiopure systems, the lattice is not perfectly hexagonal. 
Instead, a difference in nearest neighbor distances between helices of same and opposite handedness is observed, indicated by two-peak features in the RDF. 
Deconvolution of the RDF into individual EP and OP interactions shows a 2:1 ratio, which can be explained by the fact that in the line-shaped configuration each helix has two neighbors of equal handedness and four neighbors of opposite handedness. 
Because OP interactions allow for denser packing (see Section 3.1), the hexagonal lattice compresses perpendicular to the lines, enabling closer distances for OP interactions while maintaining the EP distance. 
The peak shapes of the RDF at larger distances confirm this interpretation.

Furthermore, RONN features are considered by analyzing statistical distributions of $\zeta$, $\varphi_1$, and $\chi$.
Figure~\ref{fig:RONN_RDF_OP}b shows that the sharp distribution at $\zeta = \SI{0}{\angstrom}$ for EP interactions is lost, replaced by a broad peak with maximum near \SI{0.1}{\angstrom}. 
Additionally, as indicated by the scatter plot in Figure~\ref{fig:RONN_RDF_OP}c, the defined distributions in $\varphi_1$ and $\chi$ are also lost. 
Instead, the distribution in $\varphi_1$ ranges over the entire interval from \SI{-10}{\degree} to \SI{10}{\degree} with a maximum near \SI{-8}{\degree}, and two dominant values for $\chi$ near \SI{-20}{\degree} and \SI{20}{\degree} are observed.

For OP interactions, the distribution of relative offset $\zeta$ ranges over the entire interval with a dominant peak at approximately \SI{\pm 0.75}{\angstrom}. Similar to EP interactions, the distribution of $\varphi_1$ spreads over the full range from \SI{-10}{\degree} to \SI{10}{\degree} but exhibits a less well defined peak near \SI{-7}{\degree}, while two rather sharp peaks are obtained for $\chi$ near \SI{20}{\degree} and \SI{130}{\degree}.

Comparing the obtained $\chi$ values for OP interactions with the pair interaction potential presented in Section 3.1 reveals that the system drives toward optimal OP arrangement. 
However, since this configuration is incompatible with the hexagonal lattice, frustration affects the remaining interactions, particularly EP interactions, as the interaction energy of EP configuration is approximately \SI{0.4}{\electronvolt} smaller than that of OP conformation.

Turning to the \{L$\uparrow \downarrow$-D$\uparrow \downarrow$\} system (racemic mixture with mixed up and down alignments, Figure~\ref{fig:configs_all}d), both effects reported so far appear to occur. 
On one hand, formation of parallel lines of alternating handedness is observed. 
On the other hand, as indicated by the highlighted regions in Figure~\ref{fig:configs_all}d, domains exist where right-handed helices are up-aligned while left-handed ones are oriented downwards, or vice versa. 
Preferential pairing between helices of opposite handedness and direction can be concluded from pair interaction potentials, which showed the overall most stable configuration for OA orientation. Consequently, the tendency to adopt this configuration governs the self-assembled film properties, driving the system toward the described features.

\begin{figure}[!htbp]
\centering
\includegraphics[width=1\textwidth]{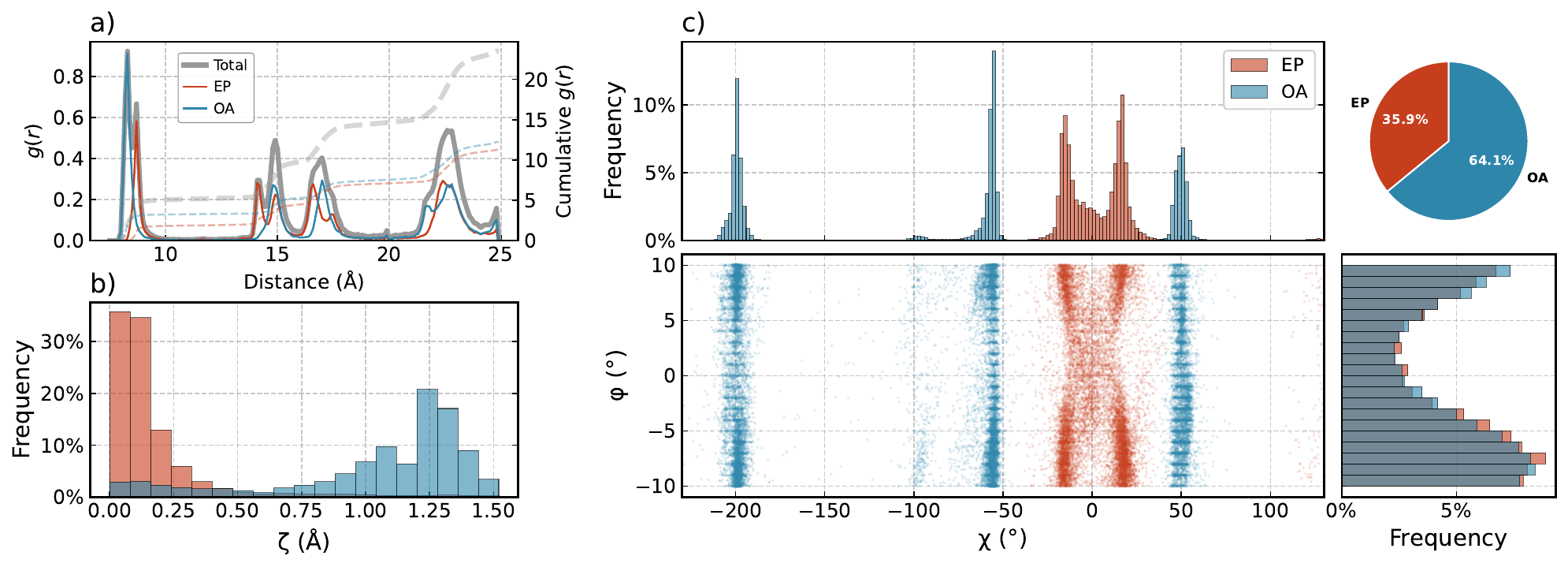}
\caption{Structural analysis of the \{L$\uparrow$-D$\downarrow$\} system. (a) Radial distribution function (RDF). (b) Distribution of the angle $\varphi_1$ for EP interactions. (c) Joint distribution of the angle difference $\chi$ and relative axial offset $\zeta$ for EP interactions. (d) Distribution of the angle $\varphi_1$ for OA interactions. (e) Joint distribution of the angle difference $\chi$ and relative axial offset $\zeta$ for OA interactions. The OA interactions show a distinct signature, confirming the drive to form optimal opposite-handed, anti-parallel pairs.}
\label{fig:RONN_RDF_OAcorr}
\end{figure}

To better understand which orientations are preferentially adopted, we considered the \{L$\uparrow$-D$\downarrow$\} system (where all right-handed helices are oriented up and left-handed ones downwards), resulting in the structure exemplarily shown in Figure~\ref{fig:configs_all}e. 
Figure~\ref{fig:RONN_RDF_OAcorr} displays the obtained characteristics for RDF and RONN. 
The complete statistical analysis of the  \{L$\uparrow \downarrow$-D$\uparrow \downarrow$\} system is presented in detail in the Supplementary Information, section~\ref{SIsection:Racemates}, where in Figure~\ref{fig:RONN_RDF_OA} the obtained RDF and RONN distribution are show.

Again, the hexagonal lattice is slightly deformed with smaller distances between helices of opposite handedness due to energetically favored OA interaction, see~\ref{fig:RONN_RDF_OAcorr}a. 
As indicated by Figures~\ref{fig:RONN_RDF_OAcorr}b and c, the EP interaction statistics reveal the same features as discussed for the \{L$\uparrow$-D$\uparrow$\} case (Figure~\ref{fig:RONN_RDF_OP}): 
a broadened $\zeta$ distribution with maximum at \SI{0}{\angstrom}, $\varphi_1$ ranging over the full interval with maximum near \SI{-8}{\degree}, and two pronounced peaks near $\chi = \SI{-10}{\degree}$ and $\chi = \SI{10}{\degree}$. 
The OA interactions distributions, however, differ significantly. 
While $\varphi_1$ follows the EP trend with a broad distribution peaking at \SI{-8}{\degree}, the $\chi$ distribution indicates three distinct peaks, while the $\zeta$ distribution spreads over the entire range with the largest contribution near \SI{-0.3}{\angstrom}.

Comparing these configurations with low-energy configurations presented in Section 3.2 reveals that self-assembled structures successfully achieve a high proportion of optimal OA pairings. 
However, geometric constraints of the hexagonal lattice introduce frustration, preventing all pairs from reaching the absolute global minimum configuration and resulting in the observed distributions for both EP and OA interaction parameters.
In conclusion, we find that the theoretical model predicts that racemic mixtures of right- and left-handed $\alpha$PA self-assemble into parallel rows of alternating handedness, giving rise to a rectangular phase.
 This result confirms the interpretation drawn from STM measurements in which the rectangular dimer phase characterized by stripe-like features was assumed to be formed by helices of opposite handedness.
The two-peak distribution in $\chi$ for EP interaction (see Figures~\ref{fig:RONN_RDF_OP}c and~\ref{fig:RONN_RDF_OAcorr}c) suggests the formation of two different angular arrangements.
Assuming, that the maximum features measured in STM experiments are not centered over the helix,  this angular difference  manifest as different distances, which are indeed observed experimentally (Figure~\ref{fig:STM}b, red line).

A direct structural interpretation of the apparent height displacement by approximately \SI{2.5}{\angstrom} within the dimer concluded from STM images is not straight forward, if one assumes parallel aligned helices with same termination groups at the surface.
It would imply either that helices are completely offset by about two alanine units (leaving a gap at the SAM-substrate interface) as proposed earlier or that helices are stretched. 
Such an offset would, however, decrease the overall interaction region, reducing the energy gain from alignment by up to \SI{150}{\milli\electronvolt} per interaction. 
A helix stretching, on the other hand, is also unlikely since this would break the intra-helix H-bonds, accounting for approximately \SI{0.4}{\electronvolt} per broken H-bond~\cite{LANTZ199961}. Both scenarios are therefore unlikely to explain the observed offset.

As our theoretical study now suggests, opposite-handed helices adopt an anti-parallel orientation, which enables formation of the most stable and densely packed configuration identified from the OA interaction potential.
Guided by these simulation results, a more plausible explanation is that the apparent height modulation originates from anti-parallel arrangement of helices bearing chemically distinct terminal groups. 
The differing electronic structures of these capping groups would naturally lead to contrast variations in STM images, thereby producing the observed height modulation without requiring physical vertical offset. 
This interpretation reconciles theoretical prediction of anti-parallel, opposite-handed dimer rows with the experimentally observed stripe-like patterns and provides a unified picture of molecular ordering within self-assembled $\alpha$-polyalanine monolayers.
\section{Conclusion}

In this study, we have developed a theoretical framework based on an effective potential derived from SCC-DFTB calculations to investigate the self-assembly of $\alpha$-polyalanine ($\alpha$PA) at the  molecular scale. 
We analyzed the generated interaction potentials and identified the specific intermolecular interactions that stabilize particular helix arrangements in isolated dimers. 
Our systematic investigation revealed that relative handedness and axial orientation of adjacent helices are the primary determinants of inter-helical packing.
We quantitatively demonstrated that opposite-handed helices in anti-parallel (OA) and parallel (OP) alignments exhibit significantly stronger binding energies and closer equilibrium distances than their equal-handed counterparts (EP, EA). This fundamental energetic preference, rooted in superior interdigitation, provides a direct atomistic rationale for Wallach's rule, explaining the denser packing observed in racemic mixtures.

By employing the generated effective potentials in heuristic optimization using simulated annealing, we successfully predicted self-assembled monolayer structures that replicate key structural motifs observed in experimental STM measurements. 
For enantiopure systems (L$\uparrow$), the dominance of EP interactions naturally leads to frustration-free, hexagonally close-packed (hcp) lattices where every neighbor pair adopts the optimal configuration. 
In contrast, racemic mixtures are driven by the superior stability of opposite-handed interactions, particularly in anti-parallel alignment, which induces structural frustration within the hexagonal lattice. 
This drives the formation of stripe-like phases with alternating chirality and rectangular structures, in excellent agreement with STM observations.

Crucially, our simulations provide a novel and more plausible interpretation of the apparent height modulation in STM images of the racemic dimer phase. 
Based on our theoretical results, we conclude that the contrast does not arise from substantial physical offset or stretching of parallel helices---scenarios that are energetically unfavorable---but rather from the anti-parallel alignment of opposite-handed helices bearing chemically distinct terminal groups. 
The differing electronic properties of these end groups at the substrate interface would naturally produce the observed height contrast in constant-current STM, reconciling experimental data with our theoretical prediction of anti-parallel, OA-stabilized dimer rows.

These findings establish robust structure-property relationships for polypeptide SAMs, directly linking chiral composition and molecular orientation to supramolecular order. 
The identified interaction motifs are critical not only for structural stability but also have profound implications for electronic and spintronic properties. 
The precise control over dipole orientation and intermolecular coupling, dictated by the identified low-energy configurations, represents a key factor modulating the Chiral-Induced Spin Selectivity (CISS) effect, explaining variations in magnetoresistance between different SAM phases.

In summary, this work moves beyond a phenomenological description to provide a predictive molecular model for chiral peptide self-assembly. 
The insights gained form a solid foundation for the rational design of peptide-based materials with tailored supramolecular order. 
Future work will integrate these structural models with charge transport calculations to explicitly unravel the mechanism of spin-selective conduction in these complex, yet elegantly ordered, biomolecular systems.

The methodological approach presented here---systematically parametrizing interactions of isolated helical dimers to construct effective potentials for SAM films---is broadly applicable and can be extended to other polypeptide systems. Of particular interest are peptides adopting non-$\alpha$-helical structures. In such systems, geometric frustration effects may emerge even in enantiopure monolayers, potentially leading to novel supramolecular architectures beyond the hexagonal and rectangular phases observed for $\alpha$PA. 
Furthermore, the generated effective potentials can be employed in kinetic Monte Carlo or molecular dynamics simulations to probe non-equilibrium kinetics of SAM formation, providing insights into nucleation, growth kinetics, and domain boundary formation.

Finally, the SCC-DFTB framework provides not only structural parameters but also complete electronic structure information for isolated helices and their pairs. This electronic foundation enables subsequent investigations into the origin of the Chiral-Induced Spin Selectivity effect, allowing direct mapping between specific molecular configurations and their spin-filtering capabilities, thereby bridging the gap between supramolecular organization and spintronic functionality.

\section*{Acknowledgements}
H.G.S, T.N.H.N, C.T.S.G, J.K acknowledge the funding trough the TRR 386 HYP*MOL of Deutsche Forschungsgesellschaft (DFG:514664767). F.G acknowledges the financial support from Conselho Nacional de Desenvolvimento Científico e Tecnológia (CNPq: 304662/2025-9) and Fundação de Amparo à Pesquisa do Estado de São Paulo (FAPESP: 2024/07315-1).

\section*{Supporting information}
S1. Structure formation during Simulated annealing,
S2. Helix-Pair Interactions of EA and OP alignments.
S3. Statistical Analysis of Low-Energy Configurations of Self-Assembled  \{L$\uparrow \downarrow$\} and \{L$\uparrow \downarrow$-D$\uparrow \downarrow$\} Films

\bibliographystyle{unsrt}
\bibliography{References}

\newcommand{\beginsupplement}{
    \setcounter{page}{1}
    \setcounter{figure}{0}
    \setcounter{table}{0}
    \renewcommand{\thepage}{S\arabic{page}}
    \renewcommand{\thefigure}{S\arabic{figure}}
    \renewcommand{\thetable}{S\arabic{table}}
    \setcounter{section}{0}
    \renewcommand{\thesection}{S\arabic{section}}
}

\clearpage
\beginsupplement

\section*{Supplementary Information}
\vspace{1cm}
\begin{center}
\textbf{\huge Simulation of Self-Assembled Monolayers of Polyalanine $\alpha$-Helices: Development and Application of an Effective Potential for Film Structure Predictions}

\vspace{1cm}

Hadis Ghodrati Saeini$^1$, Kevin Preis$^1$, Thi Ngoc Ha$^1$, Christoph Tegenkamp$^1$,  
Sibylle Gemming$^1$, Jeffrey Kelling$^{1,2}$, Florian Günther $^{3}$
\end{center}

\vspace{1cm}
\begin{enumerate}
\item Institute of Physics, Technische Universität Chemnitz, 09107 Chemnitz, Germany
\item Institute for Radiation Physics, Helmholtz-Zentrum Dresden - Rossendorf, Dresden, Germany
\item Departamento de Física, Universidade Estadual Paulista, Instituto de Geociências e Ciências Exatas, Rio Claro, Brazil
\end{enumerate}

\vspace{1em}
\section{Structure formation during Simulated annealing}
\label{sec:Waermekapazitaet}

\begin{figure}[!htbp]
\centering
\includegraphics[width=0.7\textwidth]{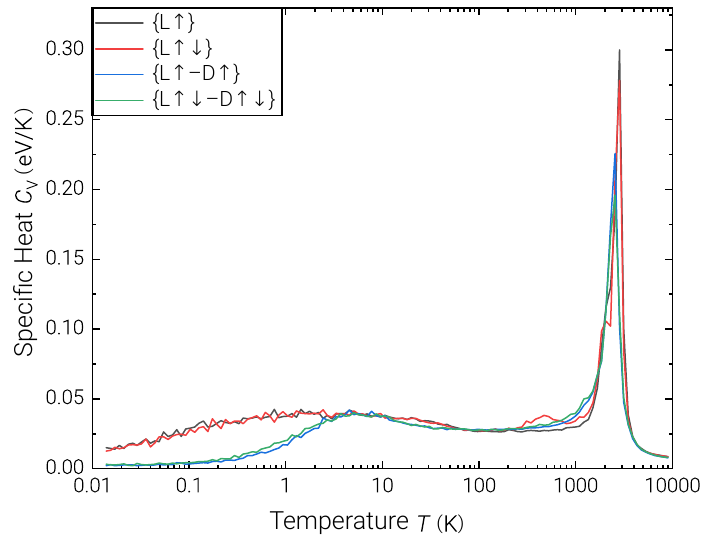}
\caption{Heat capacity $C_\mathrm{V}(T)$ for the different PA systems.}
\label{fig:HeatCap}
\end{figure}

In Figure~\ref{fig:HeatCap}, the heat capacity $C_\mathrm{V}(T)$ as a function of temperature $T$ is shown for the different ensembles. During the MC simulations, the heat capacity was evaluated from fluctuations of the internal energy. 
For a canonical ensemble, it follows from the fluctuation-dissipation theorem:
\begin{equation}
C_\mathrm{V}(T)
  = \left(\frac{\partial U}{\partial T}\right)_{V=\mathrm{const}}
  = \frac{\langle E^2 \rangle - \langle E \rangle^2}{k_\mathrm{B} T^2},
\label{eq:HeatCapLocal}
\end{equation}
where $U$ is the internal energy of the system. The ensemble averages $\langle E \rangle$ and $\langle E^2 \rangle$ are obtained from the Markov chain by simple averaging,
\[
\langle F \rangle \approx \frac{1}{N}\sum_{i=0}^{N-1} F(\vec{x}_i),
\]
with $\vec{x}_i$ denoting the generated microstates.

For simulation temperatures $T>\SI{10000}{K}$, all systems are in an amorphous state, meaning that thermal fluctuations dominate over the attractive interactions between the helices. The enantiomerically pure systems show a sharp peak in the heat capacity at approximately $T=\SI{2800}{K}$, whereas the racemic systems display a similar peak at a slightly lower temperature of about $T=\SI{2500}{K}$. These pronounced peaks identify the temperature regions where regular structures emerge and the packing density increases, indicating that attractive interactions start to outweigh thermal motion.

Around $\SI{500}{K}$, the $\{L\updownarrow\}$ systems exhibit a small additional local maximum in their heat capacity that is absent in the other systems. 
At these simulation temperature the  demixing process into enantiopure domains took place.
At simulation temperatures below $\SI{500}{K}$, all systems first show a decrease in heat capacity before another small local maximum appears between roughly $\SI1{K}$ and $\SI{20}{K}$.
For even lower temperatures, the heat capacity steadily approaches \SI0{eV K^{-1}}, with this decline occurring much faster in the racemates. 
In these systems, the increased frustration suppresses changes in the degrees of freedom of individual helices, making energy-lowering rearrangements increasingly improbable.

\section{Helix-Pair Interactions of EA and OP alignments}
\subsection{Distance-Dependent Interaction Energy}
\begin{figure}[!htbp]
\centering
\includegraphics[width=0.9\textwidth]{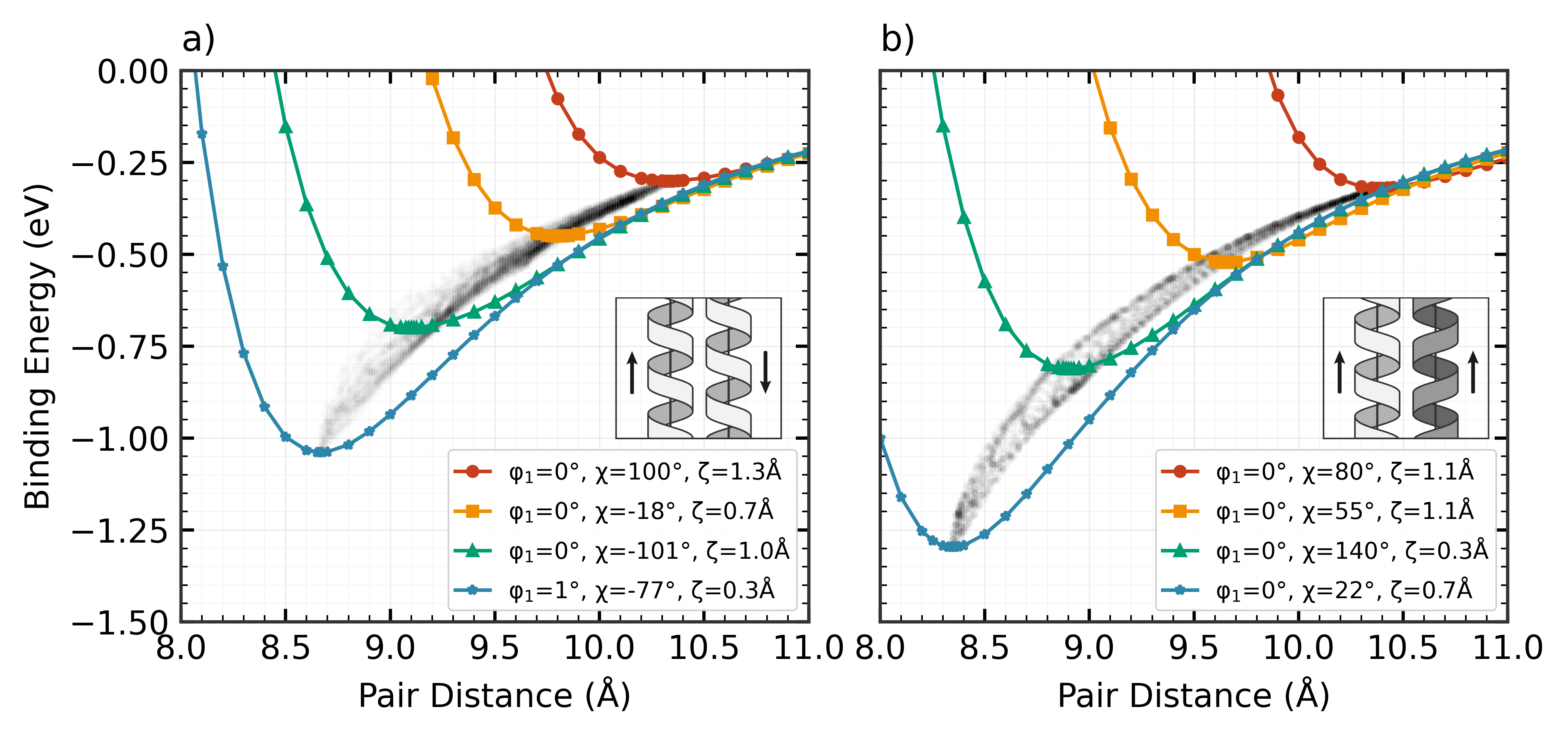}
\caption{Distance-dependent binding energy profiles for (a) EA (equal-handed, anti-parallel) and (b) OP (opposite-handed, parallel) configurations. The curves demonstrate the variation in binding strength and optimal separation distance across different relative orientations.}
\label{fig:Ropt_EA_OP}
\end{figure}

The binding energy curves and optimal distances for all simulated configurations discussed in Section~3.1 of the main text are presented here for the EA and OP cases. Figure~\ref{fig:Ropt_EA_OP}a illustrates binding energy $E_{\text{bind}}(R)$ curves for four selected relative orientations in EA alignment (colored lines), with the corresponding ($\varphi_1$,$\chi$,$\zeta$) values indicated in the legend. Figure~\ref{fig:Ropt_EA_OP}b presents corresponding results for OP alignment, where helices possess opposite handedness and are oriented parallel. The analysis reveals systematic trends in binding strength and equilibrium spacing across different orientation combinations.

\begin{figure}[!htbp]
\centering
\includegraphics[width=0.6\textwidth]{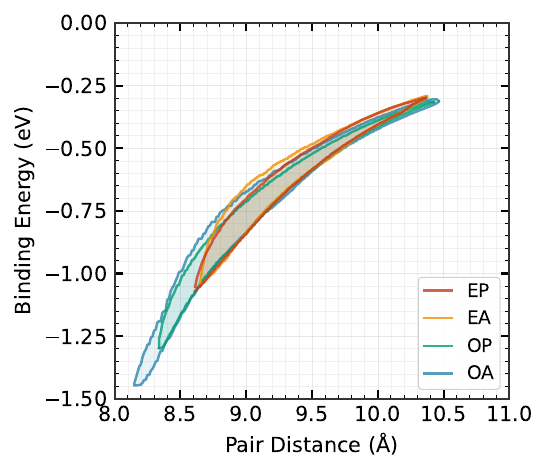}
\caption{Distribution of optimal distances for all configurations across the four interaction types (EP, EA, OP, OA).}
\label{fig:Ropt_all}
\end{figure}

To provide a comprehensive comparison of different pair interactions, Figure~\ref{fig:Ropt_all} overlays the boundary distributions of optimal distances for all possible pair configurations across the four interaction types. This overview highlights the systematic differences in packing preferences between same-handed and opposite-handed helix pairs.

\newpage

\subsection{Dependence on Relative Offset}

\begin{figure}[!htbp]
\centering
\includegraphics[width=0.8\textwidth]{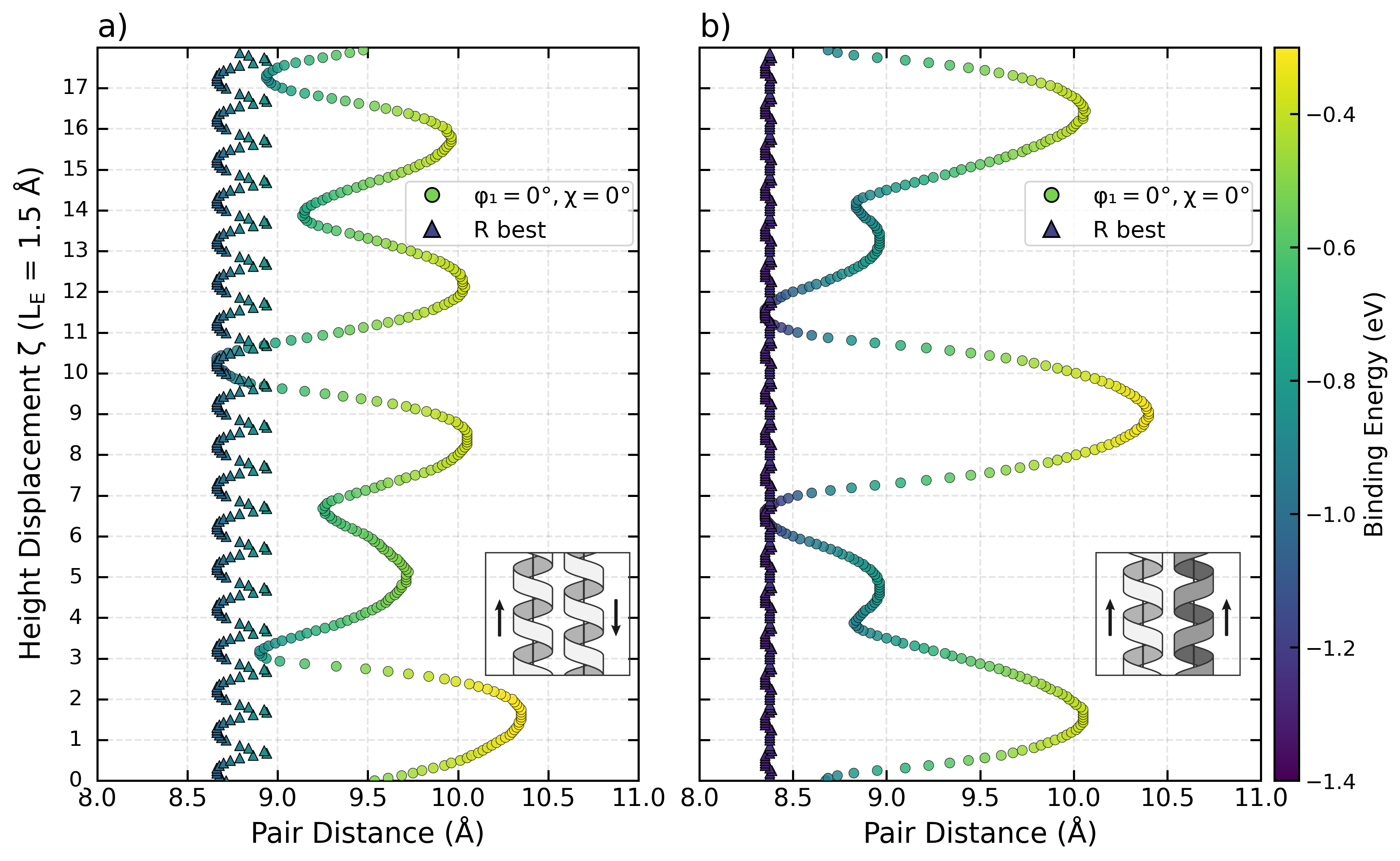}
\caption{Optimal pair distance and binding energy as a function of relative vertical offset $\zeta$ for (a) EA and (b) OP configurations. The oscillatory patterns reflect the periodic nature of helical interactions and the influence of side-chain interdigitation.}
\label{fig:zeta_EA_OP}
\end{figure}

In the main text, we investigated how the binding energy depends on the height difference and pair distances for parallel equal-handed (EP) and anti-parallel opposite-handed (OA) configurations. Figure~\ref{fig:zeta_EA_OP} extends this analysis to the EA and OP cases, showing how the relative vertical offset $\zeta$ influences the optimal inter-helical distance and binding energy for these interaction types.

\subsection{Angle Dependence}

\begin{figure}[!htbp]
\centering
\includegraphics[width=1\textwidth]{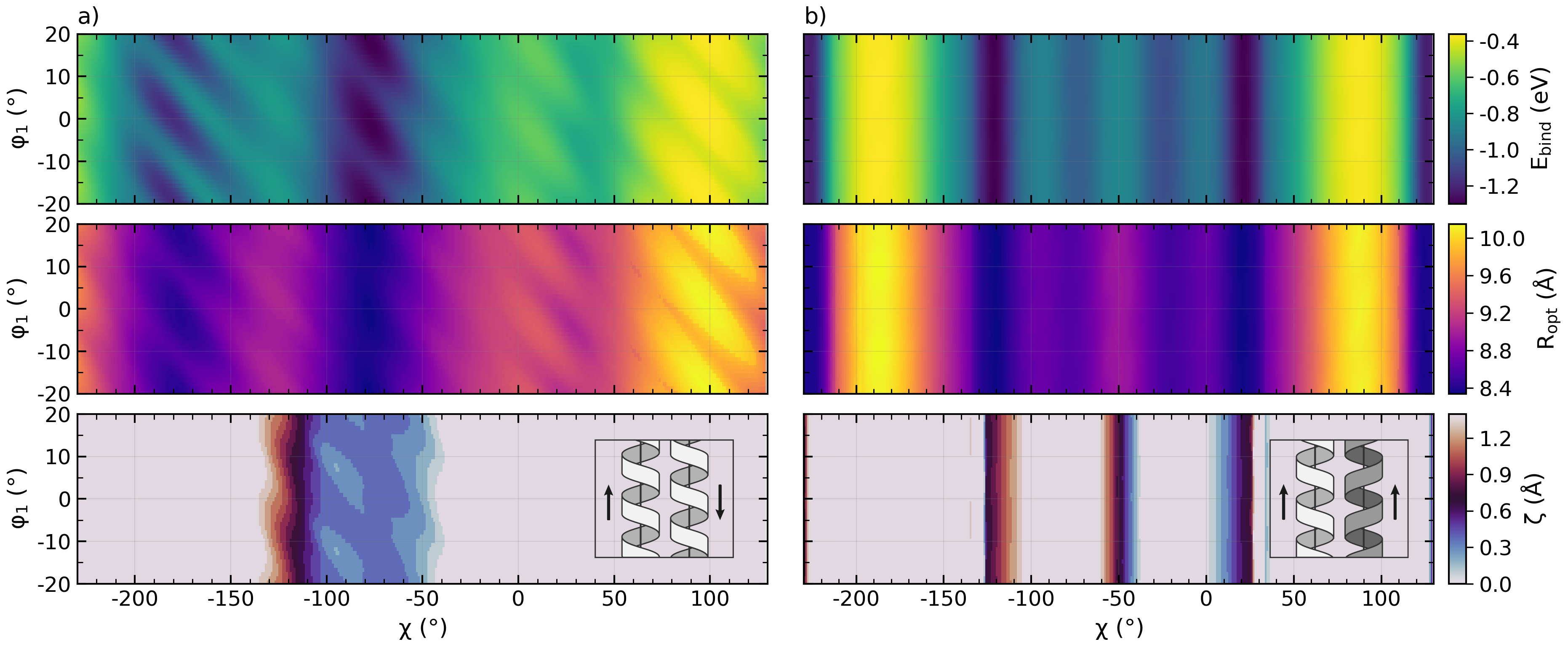}
\caption{Heatmaps of binding energy, equilibrium distance $R_{\text{opt}}$, and relative offset $\zeta_{\text{opt}}$ as functions of angular parameters $\varphi_1$ and $\chi$ for (a) EA and (b) OP configurations. }
\label{fig:heatmap_EA_OP}
\end{figure}

While the main text discussed the angular dependence for equal-handed parallel (EP) and opposite-handed anti-parallel (OA) pairs in detail, we now examine the remaining cases. Figure~\ref{fig:heatmap_EA_OP} shows the angular dependence for equal-handed anti-parallel (EA) and opposite-handed parallel (OP) configurations. For EA pairs, the binding energy landscape exhibits minima around $\varphi_1 = 0^\circ$ and $\chi \approx -70^\circ$, which can be attributed to optimal side-chain packing and hydrogen bonding patterns...
The EA configuration shows complex angular dependence while OP exhibits stripe-like features similar to OA interactions.


\subsubsection{ Local interactions for Minimum Configurations}

The global minimum configurations for EA and OP helix pairs reveal distinct structural motifs that complement the EP and OA structures discussed in the main text. For EA alignment, the optimal configuration occurs at specific angular parameters that maximize interdigitation while maintaining anti-parallel orientation. The OP global minimum configuration demonstrates how opposite-handed helices in parallel alignment achieve close packing through complementary side-chain arrangements. These structures provide insight into the diverse packing strategies available to helical polypeptides.

\begin{figure}[!htbp]
\centering
\includegraphics[width=0.8\textwidth]{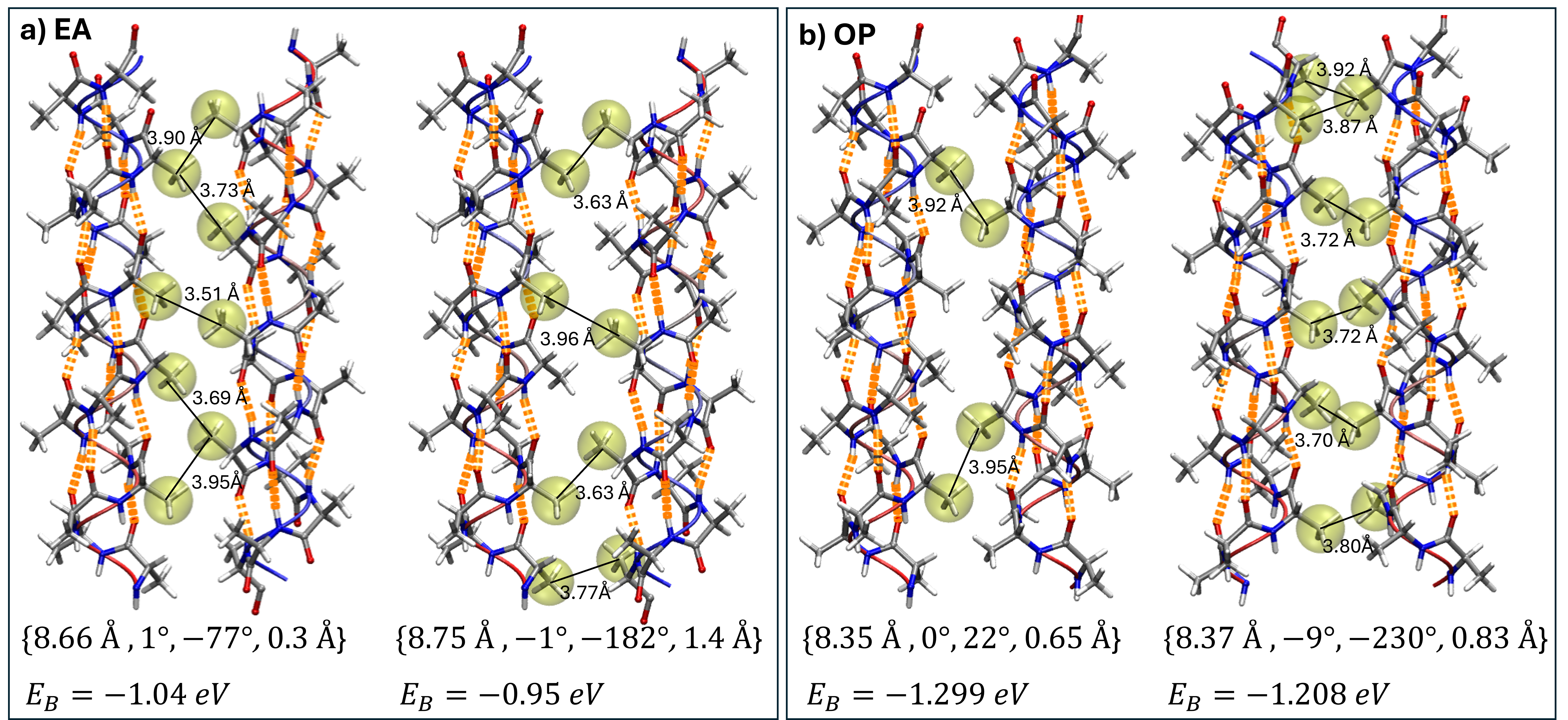}
\caption{Local Minimum configurations of EA (a) and OP (b) helix pairs, where  for better visualization the helical backbone is represented as a coil. The specific relative configuration parameters {$R$,$\varphi_1$,$\chi$,$\zeta$} as well as the corresponding binding energies $E_B$ are given below each configuration. Intra-helical hydrogen bonds stabilizing the $\alpha$ helix are indicated in orange. Methyl groups with inter-helical distances smaller than \SI4{\angstrom} are highlighted by yellow spheres with the individual distances given in the figure.}
\label{fig:SI_VMD}
\end{figure}

\section{Statistical Analysis of Low-Energy Configurations of Self-Assembled  Films}

\subsection{Structural Properties of Mixed Parallel and Anti-Parallel Enantiopure Films}

\begin{figure}[!htbp]
\centering
\includegraphics[width=1\textwidth]{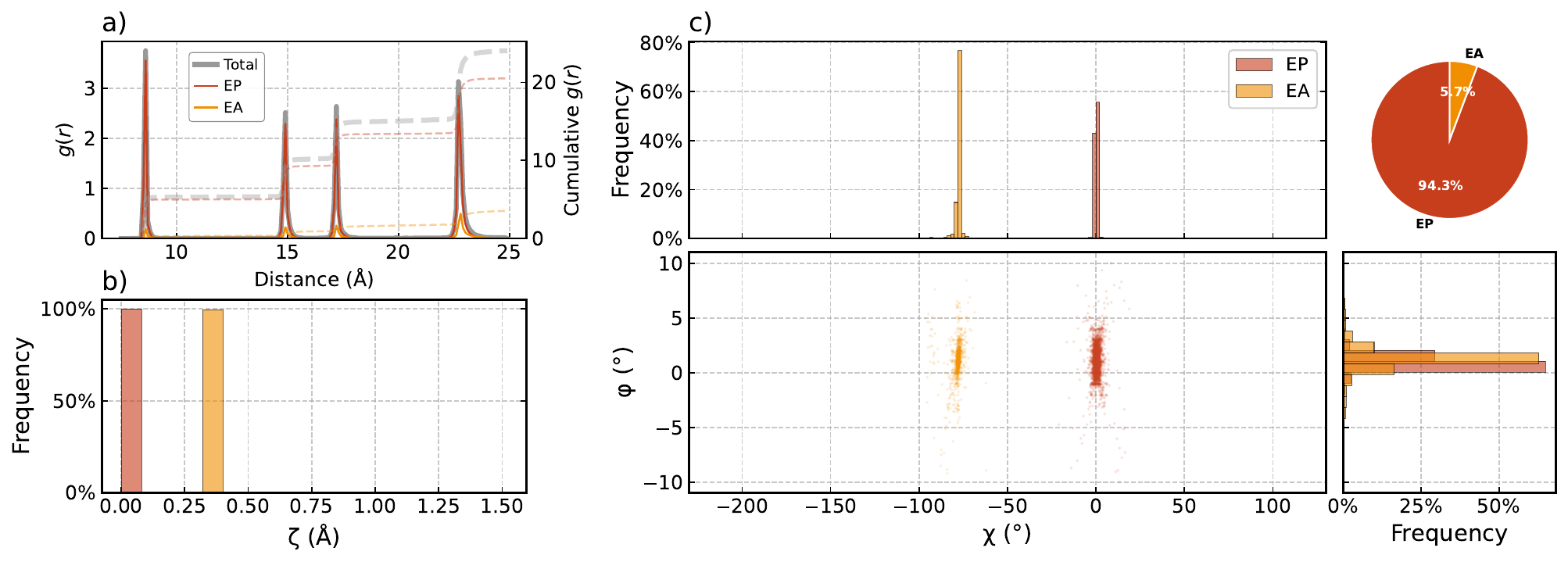}
\caption{Structural analysis of the \{L$\uparrow \downarrow$\} system showing (a) radial distribution function with domain boundary signatures, (b) distribution of relative offset $\zeta$ for EA interactions, and (c) joint distribution of angular parameters for nearest neighbors at domain boundaries.}
\label{fig:RONN_RDF_EA}
\end{figure}

As shown in Figure~\ref{fig:RONN_RDF_EA}, the \{L$\uparrow \downarrow$\} system exhibits complex domain formation with both EP and EA interactions coexisting within the same film. The radial distribution function shows characteristic features of domain boundaries, while the RONN analysis reveals how helices adapt their orientations at these interfaces. The statistical distributions provide quantitative insight into the structural frustration induced by mixed molecular orientations.

\subsection{Structural Properties of Mixed Parallel and Anti-Parallel Racemic Films}
\label{SIsection:Racemates}

\begin{figure}[!htbp]
\centering
\includegraphics[width=0.9\textwidth]{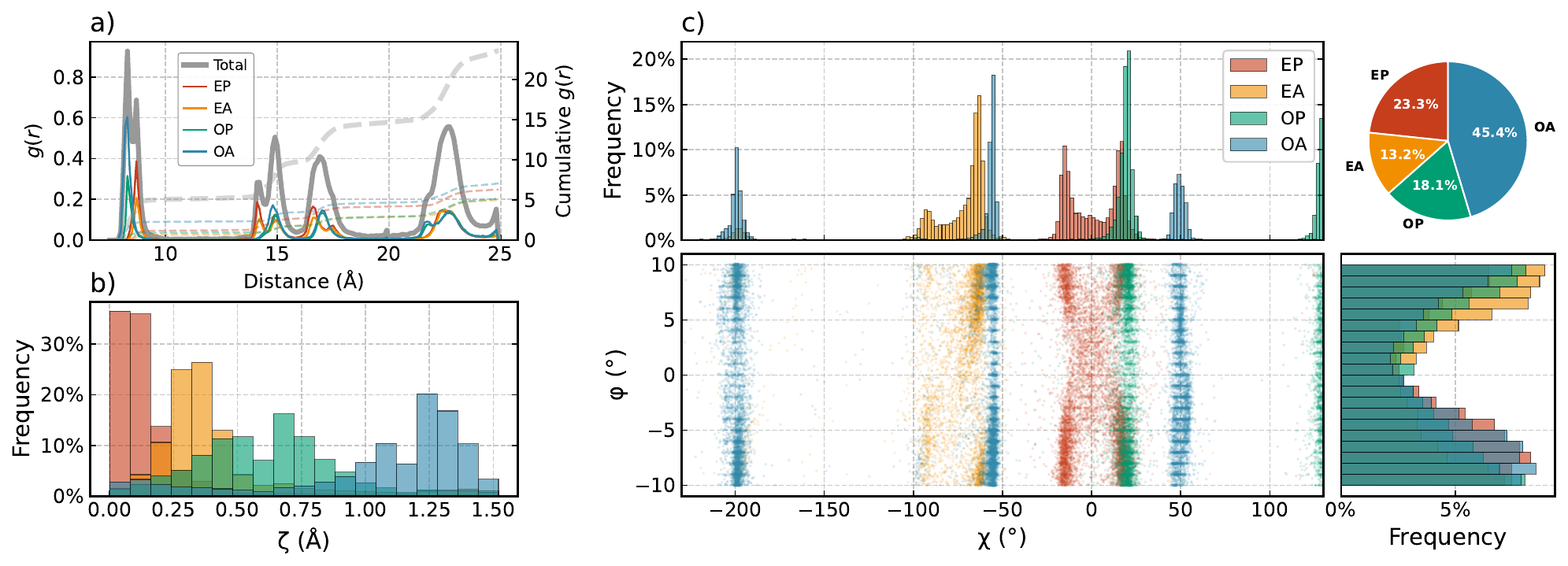}
\caption{Comprehensive structural analysis of the \{L$\uparrow \downarrow$-D$\uparrow \downarrow$\} system, including (a) radial distribution function showing complex peak structure, (b-d) RONN statistics for different interaction types.}
\label{fig:RONN_RDF_OA}
\end{figure}

The fully mixed racemic system \{L$\uparrow \downarrow$-D$\uparrow \downarrow$\} represents the most complex scenario with all four interaction types present (see Figure~\ref{fig:RONN_RDF_OA}). The structural analysis reveals how the competing preferences for OA and OP interactions lead to novel packing arrangements that differ from both pure enantiomeric and simpler racemic systems. The RDF and RONN statistics provide evidence for the formation of distinct structural domains driven by specific interaction preferences...

\end{document}